\documentclass[twocolumn]{aastex631}

\usepackage{natbib}
\usepackage{amssymb}
\usepackage{amsmath}
\mathchardef\mhyphen="2D % Define a "math hyphen"
\usepackage{verbatim}
\usepackage{indentfirst}
\usepackage{xcolor}
\usepackage{graphicx}
\usepackage{hyperref}
\usepackage{makecell}

\newcommand{\Hb}{\hb}
\newcommand{\Mbh}{{\ensuremath{\rm M_{\bullet}}}}
\newcommand{\mbh}{{\ensuremath{M_{\bullet}}}}

\newcommand{\Rnum}[1]{\uppercase\expandafter{\romannumeral #1\relax}}

\newcommand{\mgii}{{\ion{Mg}{2}}}

\newcommand{\hb}{{\rm H\ensuremath{\beta}}}

\newcommand{\Rfe}{R$_{\rm Fe}$}

\newcommand{\msun}{{\ensuremath{{M_\odot}}}}

\newcommand{\kms}{\ensuremath{\mathrm{km\,s^{-1}}}}
\newcommand{\ergs}{\ensuremath{\mathrm{erg\,s^{-1}}}}

\newcommand{\angstrom}{\textup{\AA}}

\newcommand{\eddr}{\ensuremath{\lambda_\mathrm{Edd}}}
\newcommand{\vfwhm}{\ensuremath{\mathrm{FWHM}_\mathrm{H\beta}}} % mean
\newcommand{\vfwhmmg}{\ensuremath{\mathrm{FWHM}_\text{\ion{Mg}{2}}}} % mean
\newcommand{\vsig}{\ensuremath{\mathrm{\sigma}_\mathrm{H\beta}}}
\newcommand{\vsigrms}{\ensuremath{\mathrm{\sigma}_\mathrm{H\beta, rms}}} 
\newcommand{\vsigmean}{\ensuremath{\mathrm{\sigma}_\mathrm{H\beta, mean}}}  % rms
\newcommand{\dv}{\ensuremath{\varDelta V}}
\newcommand{\lag}{\ensuremath{\tau_{\hb}}}
\newcommand{\lumcont}{\ensuremath{L_{\rm 5100}}}
\newcommand{\lumuv}{\ensuremath{L_{\rm 3000}}}

\newcommand{\lbol}{\ensuremath{L_{\mathrm{bol}}}}
\newcommand{\ledd}{\ensuremath{L_{\mathrm{Edd}}}}
\newcommand{\intscat}{\ensuremath{\sigma_\mathrm{int}}}
\newcommand{\rmsscat}{\ensuremath{\sigma_\mathrm{rms}}}
\newcommand{\uncert}[1]{\ensuremath{\varepsilon\left[#1\right]}}
\newcommand{\corr}[2]{\ensuremath{\mathrm{Corr}\left(#1,\,#2\right)}}

\newcommand{\mailto}[1]{{\normalfont\href{mailto:#1}{#1}}}

\newcommand{\snu}{\affil{Department of Physics \& Astronomy, Seoul National University, Seoul 08826, Republic of Korea; \mailto{jhwoo@snu.ac.kr}}} 
\newcommand{\texas}{\affil{Department of Physics \& Astronomy, Texas Tech University, Lubbock, TX, 79409-1051, USA}}

\begin{document}

\title{New black hole mass calibrations and the fundamental plane of the broad-line region size, luminosity, and velocity}

\author[0000-0002-8055-5465]{Jong-Hak Woo} \snu
\author[0000-0002-8055-5465]{Jimin Kim} \snu
\author[0000-0003-2010-8521]{Hojin Cho} \snu \texas
\author[0000-0002-2052-6400]{Shu Wang} \snu

\begin{abstract}
We present a new calibration of the broad-line region (BLR) size-luminosity-velocity relation using a sample of 157 AGNs with reliable \hb\ time-delay (\lag) measurements from \citet{Wang&Woo24}. By incorporating the  Eddington ratio as a third parameter, we effectively correct the systematic offset of high-Eddington AGNs in the traditional BLR size-luminosity relation.  The resulting three-parameter fit defines a fundamental plane in the 3-D space of the \lag, optical luminosity, and \hb\ velocity, with an intrinsic scatter of 0.21 dex. This tight correlation reflects the coupled effects of gas kinematics, photoionization, and BLR geometry. In turn, we develop a new method to infer \lag\ from the combination of optical luminosity and \hb\ velocity, and derive single-epoch black hole mass estimators by adopting either the full-width-at-half-maximum (FWHM) or line dispersion ($\sigma$) of the \hb\ line profile as the velocity indicator. 
The derived \lag\ shows a $\sim$0.1 dex scatter, depending on the choice of calibrations.
We show that the previous mass estimates based on the two-parameter size-luminosity relation with a 0.5 slope can be overestimated by up to 0.5 dex, demonstrating that the new mass estimator substantially changes the cosmic black hole mass density and the growth of black hole seeds in the early universe. 
\end{abstract}

%%%%%%%%%%%%%%%%%%
% Introduction   %
%%%%%%%%%%%%%%%%%%

\section{Introduction}
Accurate measurement of black hole (BH) mass is crucial for understanding the accretion physics in active galactic nuclei (AGNs) and the coevolution of massive BHs with their host galaxies. One of the most reliable methods for BH mass (\Mbh) estimation is the reverberation mapping (RM) technique \citep{Blandford+82, Peterson+93}, which measures the time delay between variability in the continuum and the broad emission lines, providing the responsivity-weighted size of the broad-line region (BLR). When combined with the BLR gas velocity from the width of broad emission lines, i.e., \Hb, the virial BH mass can be determined. Over the past two decades, the \hb\ reverberation-mapping has been applied to more than 200 AGNs, yielding reliable \hb\ lag measurements (see \citealt{Wang&Woo24} for references).

Building on these RM results, single-epoch \mbh\ estimators have been developed using the empirical relation between the \hb\ lag (\lag) and the monochromatic continuum luminosity at 5100 \AA\ (\lumcont) \citep[e.g.,][]{Kaspi+00, Vestergaard02, WooUrry02, Bentz+13}. By combining gas velocity with \lumcont\ as a proxy for the BLR size (R$_{\rm BLR}$), \mbh\ can be determined from a single-epoch spectrum albeit with larger uncertainties than reverberation-based \mbh. 

Early studies reported that R$_{\rm BLR}$ is proportional to \lumcont\ with a power of 0.7 \citep[e.g.,][]{Kaspi+00}, whereas a simple photoionization model predicts a slope of 0.5 (i.e., R$_{\rm BLR} \propto L_{\rm 5100}^{0.5}$) if the ionization parameter U and gas density is more or less constant and \lumcont\ is a good proxy for the ionizing luminosity (L$_{ion}$) since U = L$_{ion}$/(R$_{\rm BLR}^2$ n$_{e}$). After correcting for the host galaxy contribution to \lumcont, later studies reported a slope of $\sim0.5$ as expected from a simple photoionization assumption \citep{Bentz+09, Bentz+13}

The widely used size-luminosity relation with a slope of 0.5 \citep[e.g.,][]{Bentz+13} was based on a relatively small sample with a limited dynamic range. Subsequent studies reported a deviation of super-Eddington AGNs, which show a reduced R$_{\rm BLR}$ compared to that of sub-Eddington AGNs at the same luminosity \citep[e.g.,][]{Wang+14, Du+15, Grier+17}. If this is the case, then two different size-luminosity relations may be required respectively for sub- and super-Eddington AGNs \citep{Li+21}, while it is likely that the difference is caused by selection effects and the bimodal distribution of Eddington ratios. 

The relatively small size and potential bias of the reverberation-mapped AGN sample may cause systematic effect in constructing the size-luminosity relation \citep[e.g.,][]{Shen+15, Brotherton+15}.
With an enlarged and more diverse reverberation sample, \citet{Woo+24} and \citet{Wang&Woo24} updated the size-luminosity relation, finding a shallower slope ($\sim$0.4-0.42) than the  0.53-0.55 reported by \citet{Bentz+13} and a smaller scatter (i.e., rms scatter $\sim$0.3 dex).
While the AGN sample with reliable \lag\ increased by more than a factor of four over the last decade, a key change was the inclusion of high Eddington AGNs from recent studies \citep[e.g.,][]{Du+15, Du+18, Woo+24}. \citet{Woo+24} reported that the median Eddington ratio increases with luminosity in their sample, suggesting that the shallower slope is caused by the high Eddington AGNs \citep[see also discussion by][]{Du+15, Du+18}.
As the mean Eddington ratio increases with luminosity, the R$_{\rm BLR}$ shortening is more prominent at higher luminosities, flattening the slope of the size-luminosity relation.
Similar trends have been reported in the dynamical measurements of R$_{\rm BLR}$ by the GRAVITY Collaboration, 
albeit with a smaller sample size \citep{Gravity+24}.

Several scenarios have been proposed to explain the reduced BLR size in high-Eddington AGN, including slim disk self-shadowing \citep{Wang+14}, changes in the UV-optical continuum slope or BH spin \citep{Czerny+19}, wind obscuration of the far-side BLR \citep{Naddaf+25}, and changes in gas density \citep{Wu+25}. 

If the slope of the BLR size-luminosity relation is indeed flatter than the canonical 0.5, previous single-epoch \mbh\ estimates suffer a large systematic effect. In particular, high Eddington AGNs may have been overestimated in mass by factors of 2-3 when the out-dated relations were applied. 
Several studies have attempted to correct for this bias using Eddington ratio indicators, i.e., accretion rate ($\dot{m}$), the flux ratio of Fell to \hb\ lines (\Rfe), and the line profile shape \citep{Du+16, Du&Wang19}. For example, \citet{Du&Wang19} reported \Rfe\ as the best parameter among other observables to reduce the systematic offset of high Eddington AGNs, providing a three-parameter relation. Using these proxies, follow-up studies  
demonstrate that incorporating an additional parameter can improve the size-luminosity relation and single-epoch \mbh\ estimates \citep[e.g.,][]{Martinez+20, Maithil+22, Pan+25}. However, there is a large scatter as well as a systematic trend between the empirical parameters, i.e., \Rfe\ and the Eddington ratio, the predicted R$_{\rm BLR}$ and \mbh\ may suffer systematic uncertainties if the Eddington ratio is a true driver of the shortened R$_{\rm BLR}$.

In this work, we investigate the correlation of the BLR size, luminosity, and Eddington ratio using an enlarged AGN sample over a broad dynamic range in luminosity and Eddington ratios. Instead of using Eddington ratio proxies (i.e., \Rfe), we directly use the measured Eddington ratio to determine the three-parameter fit. We present new \mbh\ estimators derived  from a three-parameter fit, effectively defining a fundamental plane that accounts for the Eddington ratio effect. Section 2  describes the sample and AGN parameters. The three-parameter fitting results are presented in Section 3. followed by single-epoch \mbh\ estimators in Section 4. We discuss the implications in Section 5 and summarize our main findings in Section 6. 
Throughout this paper, we adopt a flat $\Lambda$CDM cosmology with $H_0 = 72\,{\rm km\, s^{-1}\, Mpc^{-1}}$,  $\Omega_{\rm m} = 0.3$, and $\Omega_{\rm \lambda} = 0.7$, unless specified otherwise. 

%%%%%%%%%%%%%%%%%%
% Sample         %
%%%%%%%%%%%%%%%%%%

\section{Sample and parameters}\label{s:sample}

\begin{figure}[ht!]
\centering
\includegraphics[width=0.4\textwidth]{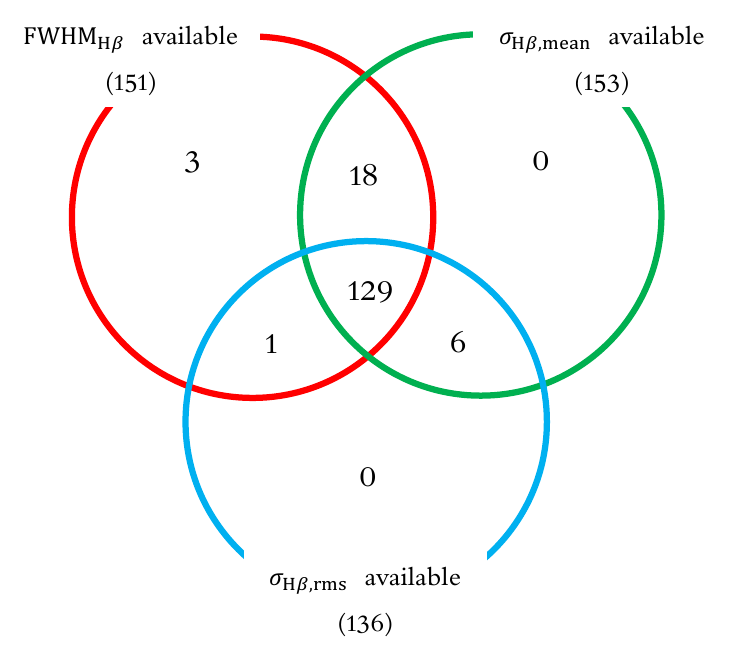}
\caption{Sample of AGNs with measured \hb\ lag from \citet{Wang&Woo24}. The number of each subsample with available \vfwhm, \vsigrms, or \vsigmean\ are presented in the circles. 
\label{fig:sample}
}
\end{figure}

We utilize the sample of AGNs with \hb-lag measurements from \cite{Wang&Woo24}, who uniformly analyzed the cross-correlation of continuum and \hb\ line light curves of 242 AGNs, using the archival data. By following the quality assessment of the lag measurements suggested by \citet{Woo+24}, they reported 157 reliable \lag\ measurements. We adopt these AGNs and their physical parameters, including \lag, \lumcont, and gas velocity.

For gas velocity, we use either the full-width-at-half-maximum (FWHM$_{\hb}$) or the line dispersion ($\sigma_{\hb}$) of the \hb\ broad line from the original papers \citep[see][]{Wang&Woo24}. Note that the \vfwhm\ measured from the mean spectrum is available for 151 AGNs, while the \vsig\ determined from either the rms spectrum or the mean spectrum is available for 136 and 153 AGNs, respectively (see Fig. 1). We will use the overlap subsample for comparing the results based on three different gas velocities.  
Through this paper, we express \lag\ in units of light-day, \lumcont\ in units of 10$^{44}$ \ergs, and gas velocity (\vfwhm\ or \vsig) in units of 1000 \kms. Note that all coefficients in the equations and tables are based on these units unless specified otherwise. 

We obtain the reverberation \mbh\ in \msun\ unit as
\begin{align}
\mbh = \frac{f}{G}\cdot c\,\lag \cdot \left(\dv{}\right)^2 \label{eq:virial_mass}, 
\end{align}
where G is the gravitational constant, and c is the speed of light, \lag\ is the lag of the \hb\ broad line (i.e.,  $c \times \lag$ = R$_{\rm BLR}$). We use \vfwhm\ or \vsig\ as gas velocity $\Delta$V, and respectively adopt the virial factor, $\log f(\vfwhm)=0.05\pm0.12$ and $\log f(\vsig)=0.65\pm0.12$ from the \mbh-stellar velocity dispersion relation calibration by \citet{Woo+15}. 
We also define Eddington ratio as
\begin{align}
\eddr = \frac{\lbol}{\ledd} = \frac{10\times (\lumcont\ \ergs)}{1.26\times (10^{38}\,\ergs) \times (\mbh/\msun)}  \label{eq:eddr}
\end{align}
using the bolometric correction of $\lbol/\lumcont = 10$ \citep{Woo&Urry02}. Note that \eddr\ is proportional to \lumcont\ and inversely proportional to \mbh, while the coefficient in Eq. 2 does not change the results in the following sections.

\section{Three-parameter fit}

In this section, we present the best-fit results by adding the third parameter to the two-parameter, \lag\ and \lumcont\ correlation (Section 3.1 and 3.2). Then, we will discuss the consistency of the results and the uncertainty of the predicted \lag\ (Section 3.3).

As presented by \citet{Woo+24}, there is a clear trend of decreasing R$_{\rm BLR}$ for higher Eddington AGNs in the size-luminosity relation. Since the size-luminosity relation is widely used to predict the BLR size and estimate \mbh, it is of importance to investigate whether the Eddington ratio effect can be corrected for by introducing a new parameter. We note that for a given luminosity, a higher Eddington AGN has a lower \mbh\ than  a lower Eddington AGN by definition. Thus, either R$_{\rm BLR}$ or $\Delta$V (or both R$_{\rm BLR}$ and $\Delta$V) should be smaller than that of a lower Eddington AGN. Consequently the 2-parameter, size-luminosity relation cannot be tight once higher-Eddington AGNs are included. In other words, the BLR size partly depends on the Eddington ratio. Therefore, it is natural to expect that a three-parameter relation may better represent the connection of the BLR properties. 

With this motivation, we introduce the third parameter to validate the three-parameter correlation as
\begin{align}
\log \lag = \alpha \log \lumcont + \beta \log X + \gamma
\label{eq:relation}
\end{align}
where $X$ determines $\tau_{\hb}$ at a fixed luminosity. 
We use \eddr\ as the third parameter to obtain the three-parameter relation among \lag, \lumcont\, and $\Delta$V in Section 3.1. 
For a consistency check we directly use $\Delta$V as the third parameter in Section 3.2.

To perform multi-parameter fitting,
we adopt a rotation-invariant regression scheme. When one parameter is clearly dependent and the other parameter is causative, an ordinary least-squares regression can be applied by minimizing the residuals along the axis of the dependent variable \citep{Isobe+90}. In contrast, when the dependency structure among variables is not clearly defined, it is well established that symmetric regression methods are more appropriate for characterizing the underlying relation \citep{Isobe+90, Akritas&Bershady96}. 

In our three-parameter fitting, it is not evident which variables should be treated as independent or dependent. Even for the two-parameter correlation between \lumcont\ and \lag, we cannot assert that the luminosity uniquely defines the time delay since the observed \lag\ strongly depends on the cloud distribution in the BLR, the projection of each cloud to the line-of-sight, the viewing angle, the opening angle, and other geometric and kinematic factors \citep[e.g.][]{Villafana+23, Naddaf+25}. In other words, the BLR response function, which governs the observed time-delay, may vary even at the same continuum luminosity. The observed \lag\ is a result of a complex interplay of gas kinematics, photoionization, and geometry.

Thus, we investigate the geometrical structure correlation without assuming cause and effect among the three parameters. To perform rotation invariant fitting, we adopt the \textsc{Python} port\footnote{\url{https://github.com/CullanHowlett/HyperFit}} of \textsc{Hyper-Fit} code \citep{HyperFit}. By minimizing orthogonal distances, this code fits an $(N-1)$-dimensional hyperplane to a set of $N$-dimensional data with heteroscedastic uncertainties. It constrains an intrinsic scatter and assumes that the uncertainties can be described as $N$-dimensional Normal distributions  Furthermore, since it fits a hyperplane to the data, the result is not affected by the choice of independent and dependent variables. Thus, this rotation-invariant \textsc{Hyper-Fit} is ideal for three-parameter fitting. 
We propagate the measurement uncertainties by considering whether the measurements are correlated or not as detailed in Appendix (\S~A.1).

For completeness, we also investigate the three-parameter correlation using conditional regression, minimizing the scatter along the axis of an assumed dependent variable. The details of this analysis are presented in Appendix (\S~A.2).

\subsection{Eddington ratio as the third parameter}

\begin{figure*}[ht!]
\centering
\includegraphics[width=0.325\textwidth]{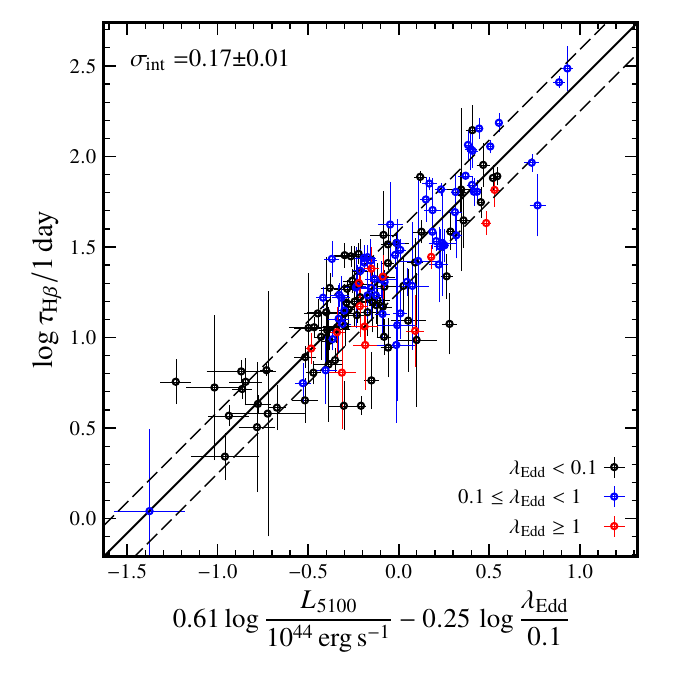}
\includegraphics[width=0.325\textwidth]{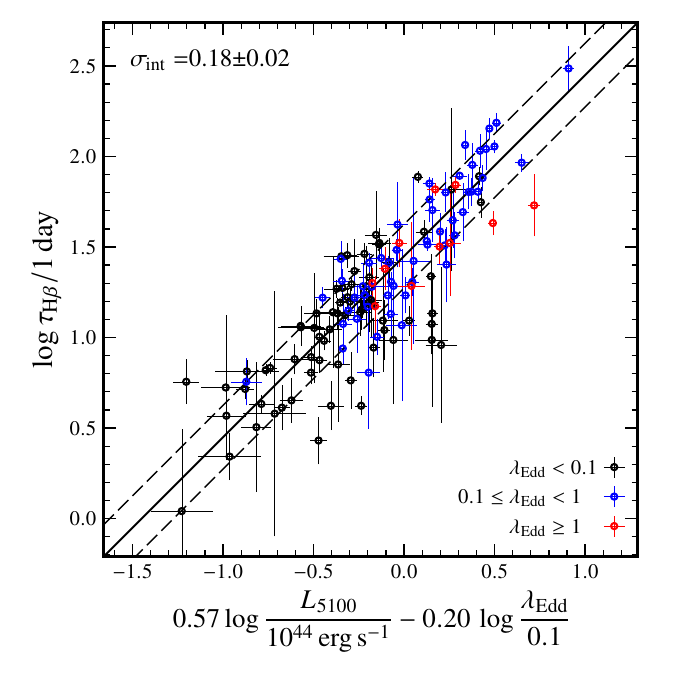}
\includegraphics[width=0.325\textwidth]{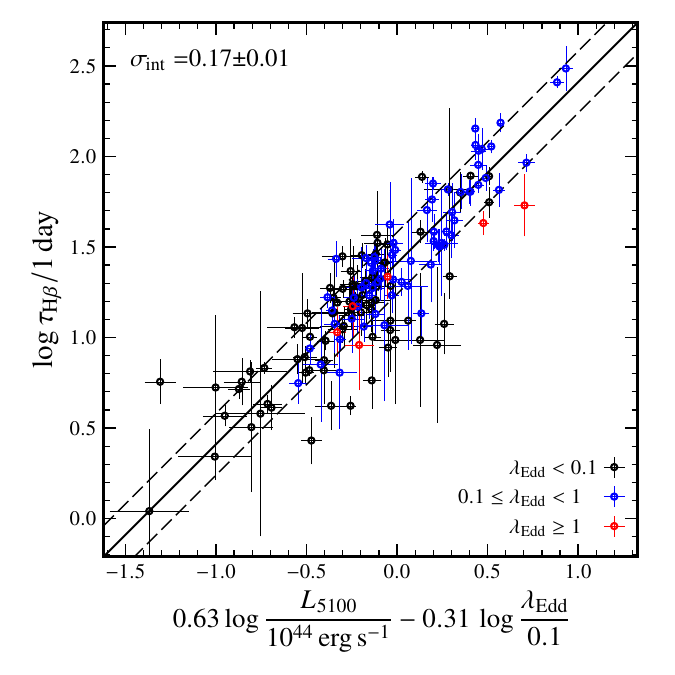}
\caption{The best-fit (solid line) with \eddr\ as the third parameter, using \vfwhm\ (left), \vsigrms\ (middle), and \vsigmean\ (right) as gas velocity. The 1-$\sigma$ range of the intrinsic scatter is denoted with dashed lines, and Eddington ratios are color-coded. We find consistent results regardless of the choice of gas velocity. 
\label{fig:EDD-fit}}
\end{figure*}

In this section we use the Eddington ratio as the third parameter (i.e., \eddr-fit).
First, using a sample of 151 AGNs, we adopt \vfwhm\ as gas velocity ($\Delta$V) in Eq. \ref{eq:virial_mass} to calculate \mbh\ and \eddr, and perform the three-parameter fitting, obtaining the best-fit relation as
\begin{align}
\begin{aligned}
%\log \lag= \left(1.42\pm 0.02\right) + \left(0.61\pm 0.03\right)\log \lumcont \\
%- \left(0.25\pm 0.03\right) \log \frac{\eddr}{0.1} 
\log \lag= \left(0.61\pm 0.03\right)\log \lumcont \\
- \left(0.25\pm 0.03\right) \log \frac{\eddr}{0.1} + \left(1.42\pm 0.02\right) 
\label{eq:bestfit_FWHM_Eddr}  
\end{aligned}
\end{align}
with an intrinsic scatter of $\intscat=0.17\pm0.01\,\mathrm{dex}$. 
The negative sign of the coefficient of \eddr\ indicates that higher Eddington AGN would have a shortened \lag\ than lower Eddington AGN (see Figure 2). Thus, the three-parameter relation reflects the correction of the \eddr\ effect present in the 2-parameter size-luminosity relation of \citet{Woo+24}. 
Since the Eddington ratio is a function of \lumcont\ and \vfwhm, we rearrange Eq.~\ref{eq:bestfit_FWHM_Eddr} by replacing \eddr\ with Eq. 1 and Eq. 2, obtaining 
\begin{align}
\begin{aligned}
\log \lag=  \left(0.48\pm 0.04\right)\log \lumcont \\
+ \left(0.68\pm 0.11\right) \log \vfwhm + \left(1.03\pm 0.08\right)
\label{eq:bestfit_FWHM_Eddr_FWHM}
\end{aligned}
\end{align}
with a propagated intrinsic scatter of $\intscat=0.21\pm0.02\,\mathrm{dex}$. 
As shown by Eq. 5, we eventually determine the three-parameter correlation among \lag, \lumcont, and \vfwhm\ from the \eddr-fit. 
Note that we present the scatter of the three-parameter fit in the \lag\ axis throughout the paper. 

Second, we use \vsigrms\ as gas velocity and perform the \eddr-fit using a sample of 136 AGNs, obtaining the best-fit as
\begin{align}
\begin{aligned}
\log \lag=  \left(0.57\pm 0.03\right)\log \lumcont \\
- \left(0.20\pm 0.04\right) \log \frac{\eddr}{0.1} + \left(1.45\pm 0.02\right)
\end{aligned}
\end{align}
with $\intscat=0.18\pm0.02\,\mathrm{dex}$. Similar to the case with \vfwhm, the correlation is tight, and the two fitting results (Eq. 4 vs. 6) are consistent, regardless of the choice of either \vfwhm\ or \vsig\ as gas velocity. 

Third, we use \vsigmean\ as gas velocity and perform the fit using a sample of 153 AGNs, obtaining the best-fit relation as
\begin{align}
\begin{aligned}
\log \lag=  \left(0.63\pm 0.03\right)\log \lumcont \\
- \left(0.31\pm 0.04\right) \log \frac{\eddr}{0.1} + \left(1.41\pm 0.02\right) 
\end{aligned}
\end{align}
with $\intscat=0.17\pm0.01\,\mathrm{dex}$. Again, the correlation is tight and consistent with the two previous cases (see Fig. 2).
We rearrange Eq. 6 and 7 into the three-parameter correlations of \lag, \lumcont, and \vfwhm\ (see Table 1). 

\renewcommand{\arraystretch}{1.5}
\begin{deluxetable*}{cllcccccc}[!ht]
\tablewidth{0.99\textwidth}
\tablecolumns{9}
\tablecaption{Three-parameter (\lag, \lumcont, and $\Delta$V) correlation \label{table:bestfit}}
\tablehead
{
    \colhead{ID}&\colhead{$X$}&\colhead{$\Delta$V}&\colhead{sample size}  &\colhead{$\alpha$}&\colhead{$\beta$}&\colhead{$\gamma$}&\colhead{\intscat}
    \\
    % \colhead{}&\colhead{}&\colhead{}&\colhead{}&\colhead{}&\colhead{}&\colhead{}
    % \\
    \colhead{(1)} &\colhead{(2)} &\colhead{(3)}&\colhead{(4)}&\colhead{(5)}&\colhead{(6)}&\colhead{(7)}&\colhead{(8)} 
}
\startdata
\textbf{Case 1}& \textbf{\boldmath $\displaystyle\eddr$-fit}   & \textbf{\boldmath \vfwhm} & \textbf{151} & \textbf{\boldmath $0.48\pm 0.04$} & \textbf{\boldmath $ 0.68\pm 0.11$} & \textbf{\boldmath $1.03\pm 0.08$} & \textbf{\boldmath$0.21\pm0.02$} \\
\textbf{Case 2}& \textbf{\boldmath $\displaystyle\eddr$-fit}    & \textbf{\boldmath \vsigrms}     & \textbf{136} & \textbf{\boldmath $0.46\pm 0.05$} & \textbf{\boldmath $ 0.50\pm0.12$} & \textbf{\boldmath $1.32\pm 0.05$} & \textbf{\boldmath$0.21\pm0.02$} \\
\textbf{Case 3}& \textbf{\boldmath $\displaystyle\eddr$-fit}    & \textbf{\boldmath \vsigmean}     & \textbf{153} & \textbf{\boldmath $0.47\pm 0.05$} & \textbf{\boldmath $ 0.89\pm0.15$} & \textbf{\boldmath $1.17\pm 0.07$} & \textbf{\boldmath$0.21\pm0.02$} \\
\hline
Case 4& $\displaystyle\vfwhm$-fit      & \vfwhm & 151 & $0.46\pm 0.02$ & $ 0.42\pm 0.11$ & $1.18\pm 0.06$ & $0.20\pm0.02$ \\
Case 5&$\displaystyle\vsigrms$-fit    & \vsigrms & 136  & $0.45\pm 0.03$ & $ -0.48\pm 0.17$ & $1.52\pm 0.04$ & $0.19\pm0.02$ \\
Case 6&$\displaystyle\vsigmean$-fit & \vsigmean & 153 & $0.44\pm 0.04$ & $ -1.76\pm 0.94$ & $1.85\pm 0.24$ & $0.39\pm0.11$ 
\enddata
\tablecomments{Column (1): Case number. Column (2): The third parameter used in Eq.~\ref{eq:relation}. Column (3): Choice of gas velocity. Column (4): Sample size. Columns (5): Best-fit coefficient of \lumcont\ in Eq.~\ref{eq:relation}. Columns (6) : Best-fit coefficient of velocity (either \vfwhm\ or \vsig\ depending on the choice of $\Delta$V) in Eq.~3. Columns (7): Best-fit coefficient in Eq.~\ref{eq:relation}. Columns (8): Intrinsic scatter of the relation. Note that the representative values are the maximum posterior estimators, and the uncertainties are the central 68\% intervals. $\log f(\vfwhm)=0.05\pm0.12$ were adopted from \citet{Woo+15} to obtain relevant fits.}
\end{deluxetable*}

\subsection{Gas velocity as the third parameter}

\begin{figure*}[ht!]
\centering
\includegraphics[width=0.325\textwidth]{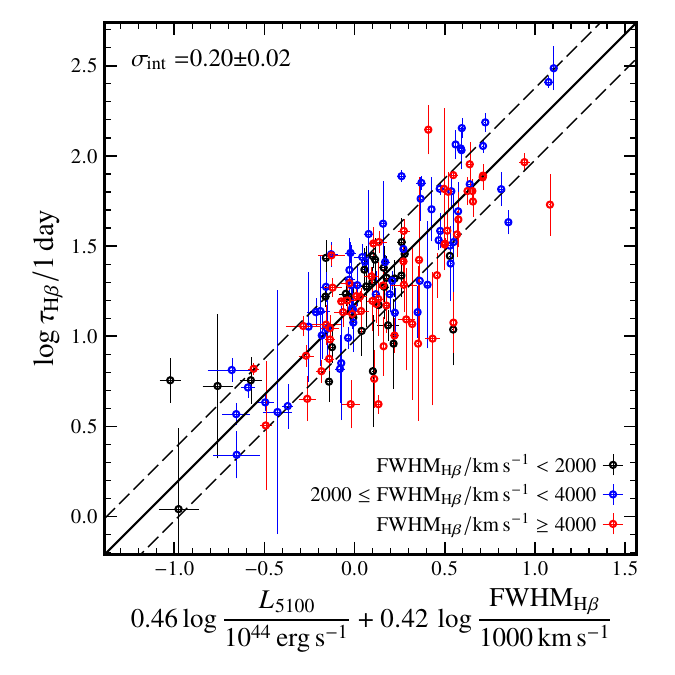}
\includegraphics[width=0.325\textwidth]{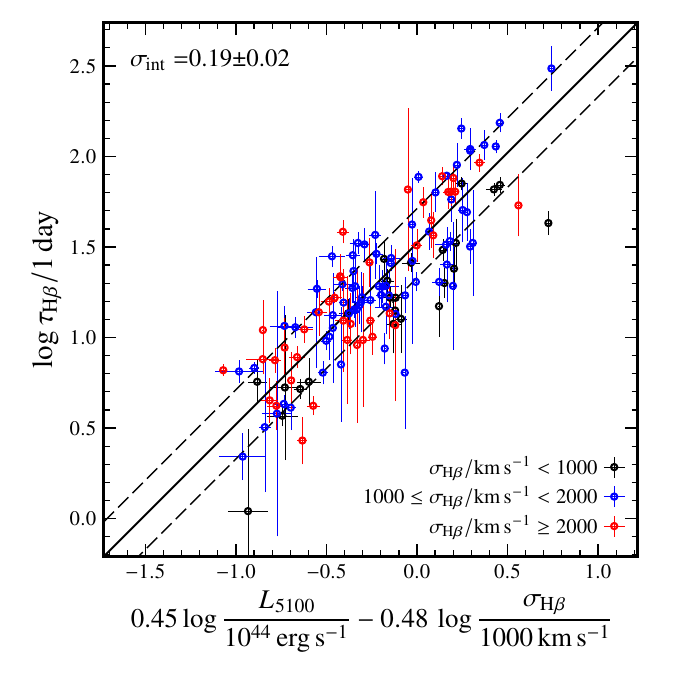}
\includegraphics[width=0.325\textwidth]{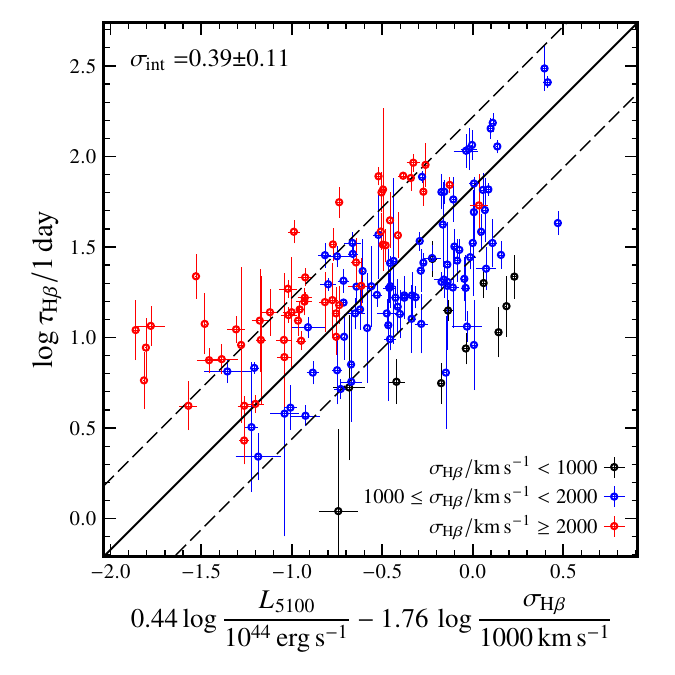}
\caption{The best-fit (solid line) with gas velocity as the third parameter, using \vfwhm\ (left), \vsigrms\ (middle), and \vsigmean\ (right) as gas velocity. The 1-$\sigma$ range of the intrinsic scatter is denoted with dashed lines, and Eddington ratios are color-coded. We find an opposite sign of the coefficients of gas velocity, depending on the choice of gas velocity. In contrast to the \eddr-fit, we find relatively large intrinsic scatter. 
\label{VEL-fit}}
\end{figure*}

\begin{figure*}[ht!]
\centering
\includegraphics[width=0.32\textwidth]{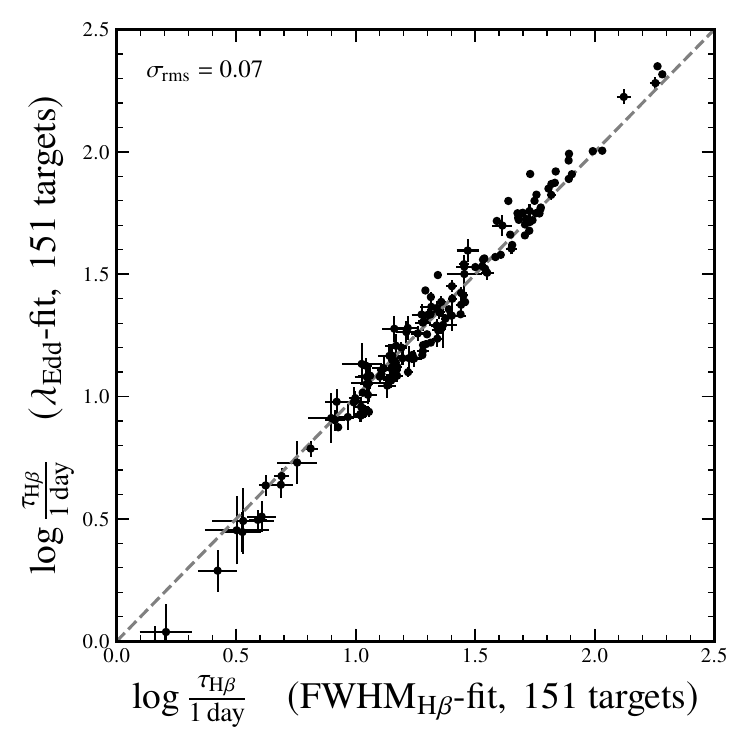}
\includegraphics[width=0.32\textwidth]{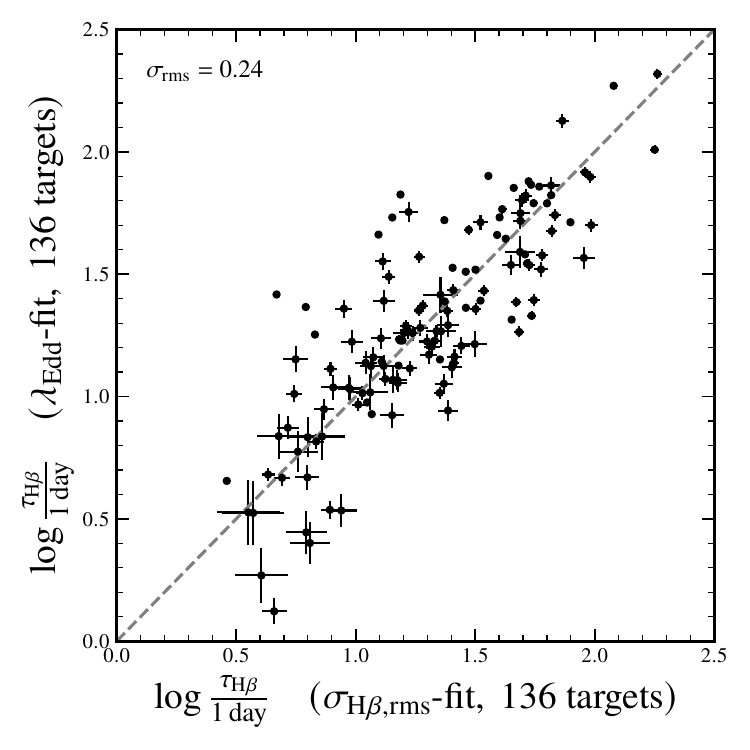}
\includegraphics[width=0.32\textwidth]{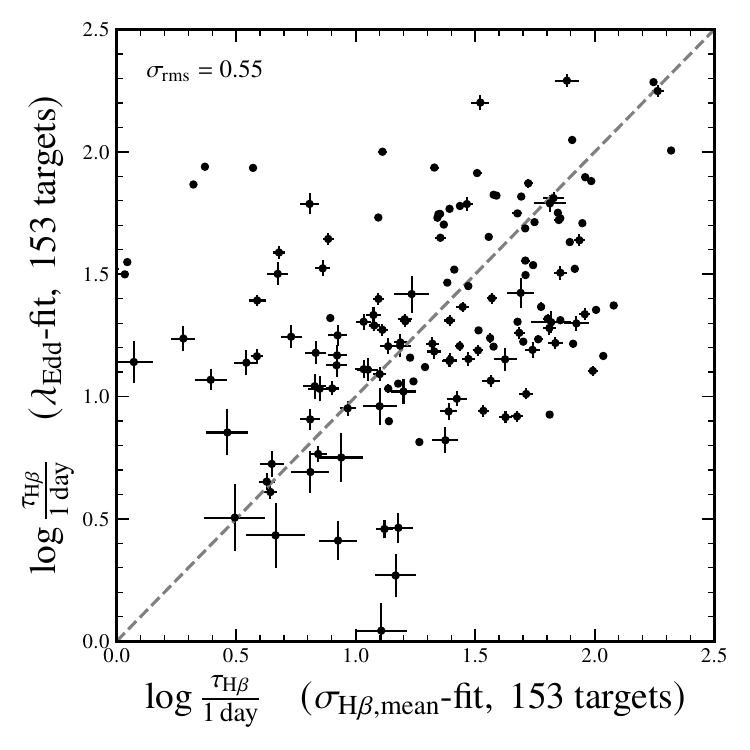}
\caption{Left: Comparison of the expected \lag\ based on either the \eddr-fit or the \vfwhm-fit using \vfwhm\ as gas velocity. 
Middle: Comparison of the expected \lag\ based on either the \eddr-fit or the \vsigrms-fit using \vsigrms\ as gas velocity. 
Right: Comparison of the expected \lag\ based on either the \eddr-fit or the \vsigmean-fit using \vsigmean\ as gas velocity. 
}
\end{figure*}

For consistency check, we use gas velocity as the third parameter and perform the three-parameter fit. Similar to Section 3.1, we adopt three gas velocities, i.e., \vfwhm, \vsigrms, and \vsigmean, respectively, in Eq. \ref{eq:virial_mass} to calculate \mbh\ and \eddr. 
Using \vfwhm\ as the third parameter (\vfwhm-fit), we obtain the best-fit relation as
\begin{align}
\begin{aligned}
\log \lag=  \left(0.46\pm 0.02\right)\log \lumcont \\
+ \left(0.42\pm 0.11\right) \log \vfwhm + \left(1.18\pm 0.06\right) 
\label{eq:bestfit_FWHM}
\end{aligned}
\end{align}
with an intrinsic scatter of $\intscat=0.20\pm0.02\,\mathrm{dex}$.
% (see Figure~\ref{fig:RL_FWHM}).

In the case of the \vsigrms-fit, we obtain
\begin{align}
\begin{aligned}
\log \lag = \left(0.45\pm 0.03\right)\log \lumcont \\
- \left(0.48\pm 0.17\right) \log \vsig +  \left(1.53\pm 0.04\right)  
\end{aligned}
\end{align}
with $\intscat=0.19\pm0.02\,\mathrm{dex}$. While the intrinsic scatter is small, similar to the case of the \eddr-fit, we obtain an opposite sign of the velocity coefficient from these two cases (Eq. 8 vs. 9, see Figure 3). We will discuss this inconsistency in the next section. 

From the \vsigmean-fit, we obtain the best-fit relation as
\begin{align}
\begin{aligned}
\log \lag = \left(0.44\pm 0.04\right)\log \lumcont \\
- \left(1.76\pm 0.94\right) \log \vsig + \left(1.85\pm 0.24\right)  
\end{aligned}
\end{align}
with $\intscat=0.39\pm0.11\,\mathrm{dex}$. We find a substantially large intrinsic scatter, in contrast to the case of the \eddr-fit or \vfwhm-fit and \vsigrms-fit (see Figure 3). 

In contrast to the \eddr-fit, we find the results with gas velocity as the third parameter are not reliable as the signs of the coefficient of the third parameter depends on the choice of gas velocity (\vfwhm\ vs. \vsig) and the scatter is substantially large in the case of the \vsigmean-fit. These results indicate that by adding the gas velocity as the third parameter, we obtain no better relation than the two-parameter size-luminosity relation (for comparison, see Table 1). 

In Figure 4, we compare the predicted \lag\ based on the \eddr-fit or \vfwhm-fit relation (i.e., Eq. 4 or 8) using a pair of \lumcont\ and \vfwhm\ of each AGN in the sample. We find negligible difference (i.e.,  $\rmsscat=0.07\,\mathrm{dex}$) when we choose \vfwhm\ as gas velocity. 
In contrast, when \vsigrms\ is used as gas velocity, we find a significant scatter of $\rmsscat = 0.24\,\mathrm{dex}$ between the predicted \lag\ from the \eddr-fit (Eq. 6) and the \vsigrms-fit (Eq. 9). This large scatter is caused by the opposite signs of the velocity coefficient in the three-parameter relations (Case 2.  vs. Case 5 in Table 1). However, the predicted \lag\ based on the \eddr-fit seems reliable since the \eddr-fit shows consistency regardless of the choice of gas velocity as discussed in Section 3.1. Thus, we interpret that the scatter between the two predicted \lag\ is mainly caused by the systematic uncertainties of the \vsigrms-fit results. 
When we choose \vsigmean\ as gas velocity, we obtain an even larger scatter of $\rmsscat = 0.55\,\mathrm{dex}$ between the predicted \lag\ from the \eddr-fit (Eq. 6) and the \vsigmean-fit (Eq. 10). This inconsistency is partly expected from the large scatter in the three-parameter relation with \vsigmean\ as shown in Fig. 3 (right).
These comparison results suggest that \lag\ is not reliably predicted when gas velocity is used as the third parameter.

 \subsection{Fundamental plane of \lag, \lumcont, and \vfwhm}
 
 \begin{figure}[ht!]
\centering
\includegraphics[width=0.45\textwidth]{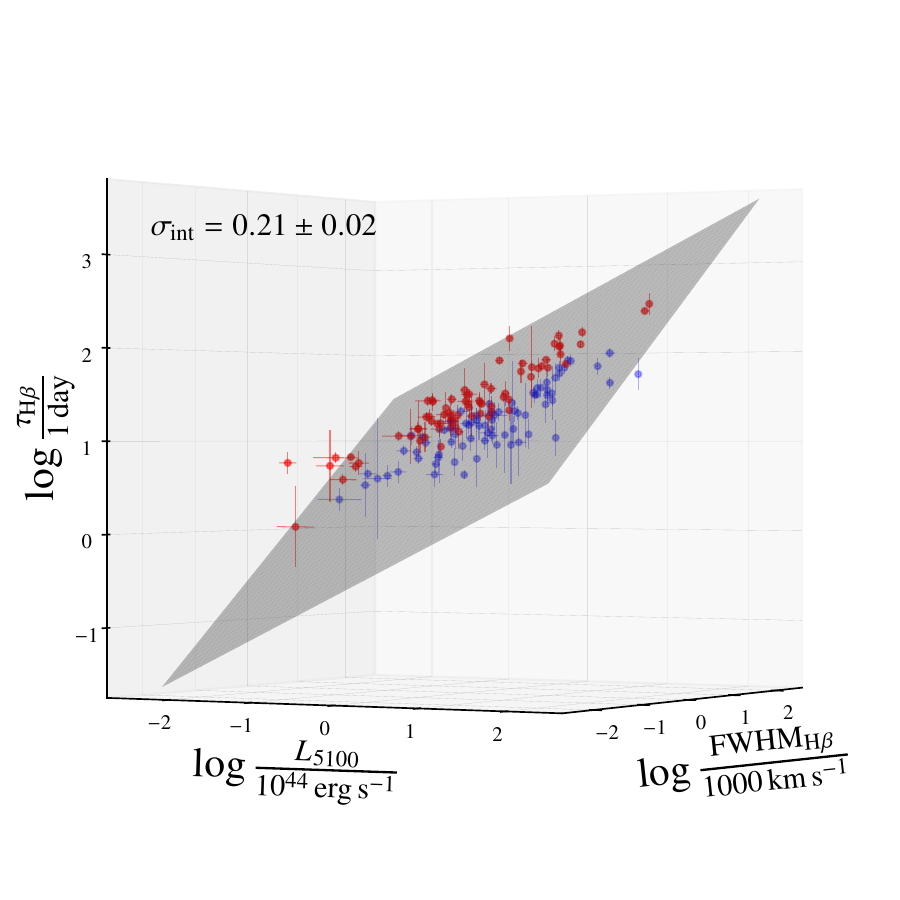}
\caption{The fundamental plane defined in the 3-D space of \lag, \lumcont, and \vfwhm. We denote the objects above and below the plane with red and blue symbols, respectively.
\label{fig:fundamental}}
\end{figure}

The \eddr-fit results indicate that the correlation of \lag, \lumcont, and \vfwhm\ (or \vsig) is tight, 
providing a new calibration for determining \lag, after effectively correcting for the influence of Eddington ratio.
In Figure 5 we present the fundamental plane of \lag, \lumcont, and \vfwhm\ relation as derived from Eq. 5. As expected, the distribution of AGNs in the 3-d space is not random. Instead, a fundamental plane is empirically defined. The tilt of the plane defined by the coefficient of velocity in Eq. 5 indicates that Eddington ratio varies along the plane. For example, at fixed luminosity, \lag\ and \vfwhm\ show a positive trend, indicating that velocity becomes smaller when the BLR size becomes smaller, thus Eddington ratio becomes larger. Therefore, the tilt of the fundamental plane is defined for correcting for the Eddington ratio effect. Note that the tilt of the fundamental plane is consistent regardless of the choice of the gas velocity (\vfwhm\ or \vsig) and the sample (151, 136, or 153 AGNs) when we choose \eddr\ as the third parameter. 

In contrast, when we use gas velocity as the third parameter, we find discrepancy between the case of \vfwhm\ and \vsig. While the \vfwhm-fit provides a consistent tilt of the plane (i.e., a positive sign of the velocity term in the three-parameter relation), the \vsigrms-fit and \vsigmean-fit show an opposite tilt of the plane (see Table 1). It seems that the \vsig-fit is less reliable since the dynamic range is limited to $\sim1$ dex compared to the \eddr-fit.

Since the motivation of introducing the third parameter is to correct for the effect of the Eddington ratio, we expect a negative sign of the \eddr\ coefficient (i.e., a positive sign of the velocity term). In fact, we find the \eddr-fit always shows a positive sign of the velocity coefficient as expected. Thus, we select the \eddr-fit (Case 1, Case 2, and Case 3 depending on the choice of gas velocity) as our recommendation based on the theoretical and empirical consistency.

%%%%%%%%%%%%%%%%%%
% SE Mass        %
%%%%%%%%%%%%%%%%%%
\section{Single-Epoch Mass Estimators}

In this section, we present new single-epoch mass estimators by combining the virial mass equation (Eq. 1) and the fundamental plane presented in Section 3. We first provide optical mass estimators with \lumcont\ and \hb\ velocity in Section 4.1. 
Then, we present UV mass estimators with the continuum luminosity measured at 3000 \AA\ (\lumuv) and the line width of the \mgii\ broad line in Section 4.2.  

\subsection{H$\beta$ Mass Estimators}

We express the single-epoch \mbh\ as
\begin{align}
\begin{aligned}
\log \mbh  =  p \log \lumcont + q \log \Delta V + C_{M},
\end{aligned}
\label{eq:singleMBH}
\end{align}
and derive the coefficients by combining Eq. 1 and the \eddr-fit results from Section 3. 
First, we use \vfwhm\ as $\Delta$V in Eq.~\ref{eq:virial_mass}, and adopt Eq. 5 (i.e., Case 1 in Table 1) to calibrate \mbh\ as 
\begin{align}
\begin{aligned}
\log \mbh  = 6.37\pm0.22 +\left(0.48\pm 0.04\right)\log \lumcont \\
+ \left(2.68\pm 0.11\right) \log \vfwhm
\end{aligned}
\label{eq:hb_se_vfwhm}
\end{align}
with an intrinsic scatter of $\intscat=0.21\pm0.02\,\mathrm{dex}$.
Second, we derive the \vsigrms-based mass estimator from the \eddr-fit (Case 2) by 
adopting \vsigrms\ as $\Delta$V, as
\begin{align}
\begin{aligned}
\log \mbh  = 7.26\pm0.21 +\left(0.46\pm 0.05\right)\log \lumcont \\
+ \left(2.51\pm 0.13\right) \log \vsigrms
\end{aligned}
\label{eq:hb_se_vfwhm}
\end{align}
with an intrinsic scatter of $\intscat=0.21\pm0.02\,\mathrm{dex}$.
Third, we adopt the result from the \eddr-fit with \vsigmean\ as $\Delta$V (Case 3), obtaining 
\begin{align}
\begin{aligned}
\log \mbh  = 7.11\pm0.14 +\left(0.47\pm 0.05\right)\log \lumcont \\
+ \left(2.89\pm 0.15\right) \log \vsigmean
\end{aligned}
\label{eq:hb_se_vfwhm}
\end{align}
with an intrinsic scatter of $\intscat=0.21\pm0.02\,\mathrm{dex}$.
Note that the coefficient of the velocity term is not a factor of 2 as expected from the virial relation. However, this does not mean inconsistency with the virial relation since the BLR size varies partly depending on the velocity as shown by the fundamental plane. 

Depending on the availability of the line width measurements, users can choose one of the mass estimators. 
If only a single-epoch spectrum is available, the FWHM or line dispersion of \hb\ can be measured and applied to 
Eq.~12 or Eq.~14, assuming that the mean spectrum is representative of single-epoch spectra. 
It is well known that line-width measurements from the mean and rms spectra are systematically different. 
For example, \citet{Collin06} compared mean and rms line profiles for a set of reverberation-mapped AGNs and found that the 
\hb\ line width measured from the mean spectrum is typically broader by $\sim20\%$ than that from the rms spectrum \citep[see also][]{Wang+19}.
The difference is presumably caused by the fact that the rms spectrum represents the variable part of the BLR emission, 
while the mean spectrum includes additional non-variable wing components. For the reverberation-mapped AGNs used in this study, we report that the line dispersion from mean spectra is broader by $\sim$20\% than that from rms spectra (i.e., $\sigma_{\rm \hb, mean}$/$\sigma_{\rm \hb, rms}$ = $1.21\pm0.33$).

For a consistency check, we compare the inferred \lag\ and \mbh\ using either \vfwhm\ or \vsigrms\ as gas velocity in Figure 6. First, we find a consistency of the predicted \lag\ with $\rmsscat < 0.1$. In contrast, we find a 0.39 dex scatter between the two mass estimates, indicating single-epoch \mbh\ estimates can be somewhat different depending on the choice of gas velocity.
However, this scatter is not mainly due to the uncertainty of the predicted \lag\ since the predicted \lag\ is consistent within 0.1 dex regardless of the choice of the calibrations. In fact, the scatter in the mass comparison is mainly caused by the difference of the two line width measurements. As shown in Figure 6 (right panel), we find an intrinsic scatter of 0.13 dex between \vfwhm\ and \vsigrms, which is translated into a \mbh\ scatter of 0.33-0.34 dex since \mbh\ $\propto \Delta$V$^{2.6-2.7}$ in Eq. 12 and 13. Thus, the scatter in the mass comparison can be almost entirely due to the intrinsic scatter between the two line width measurements. Note that this 0.39 dex scatter can be considered as a systematic uncertainty of both RM-based \mbh\ and single-epoch \mbh, depending on the choice of \vfwhm\ or \vsigrms. For a given velocity measure (either \vfwhm\ or \vsigrms), however, the uncertainty of the single-epoch \mbh\ caused by the prediction of \lag\ is 0.1 dex. 

In Figure 7, we also compare the predicted \lag\ and \mbh\ using either \vfwhm\ or \vsigmean. The inferred \lag\ is consistent regardless of the choice of the three-parameter correlations. However, we find an intrinsic scatter of 0.33 dex, which is mainly caused by the scatter between \vfwhm\ and \vsigmean. Note that the ratio of \vfwhm/\vsigmean\ is $2.02\pm0.57$, indicating a substantially large range, but the range of the ratio is slightly smaller than the case of \vfwhm/\vsigrms\ ($2.49\pm0.77$). We suspect that there are systematic uncertainness in the measurements of line widths since various groups applied slightly different methods to their \hb-lag AGN sample \citep[see references in][]{Wang&Woo24}.

Based on these considerations, we conclude that the new \mbh\ estimators are reliable since the predicted \lag\ does not depend on the calibration with \vfwhm\ or \vsig. On the other hand, not only single-epoch mass but also the reverberation-based mass suffers the systematic uncertainty depending on the choice of gas velocity (either \vfwhm\ or \vsig).

To investigate the systematic change of the single-epoch mass estimates, we adopt the \vfwhm\ and \lumcont\ of $\sim$14,000 type 1 AGNs from \citet{Rakshit+20} and calculate \mbh\ using Eq. 12. For the previous mass estimate, we adopt the size-luminosity relation from \citep[e.g.,][]{Bentz+13}. In Figure 8 we compare the two mass estimates, finding a clear systematic offset. For high Eddington AGN, the previous mass was overestimated by up to 0.5 dex compared to the new mass estimate, indicating that recalibration of \mbh\ is crucial for high Eddington AGNs. 

\begin{figure*}[ht!]
\centering
\includegraphics[width=0.30\textwidth]{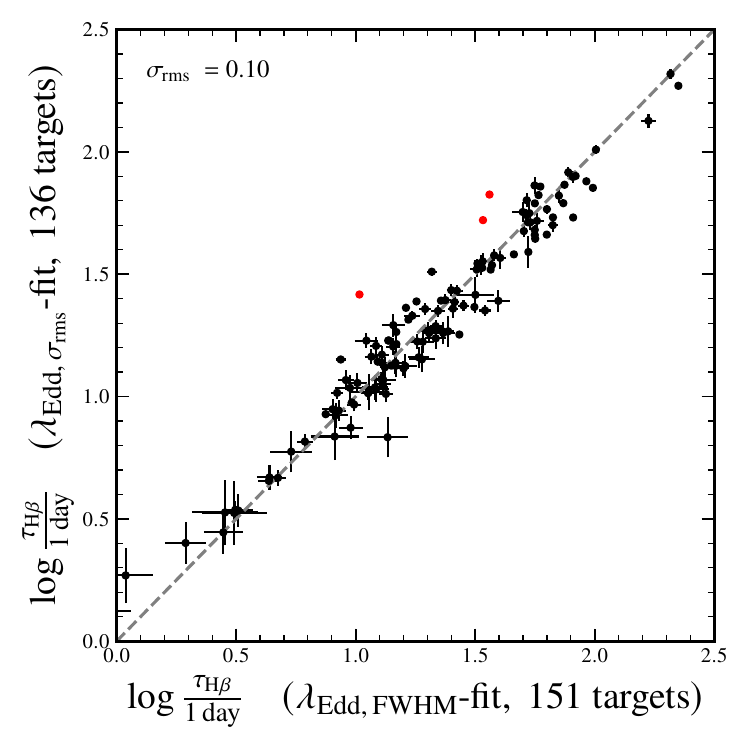}
\includegraphics[width=0.30\textwidth]{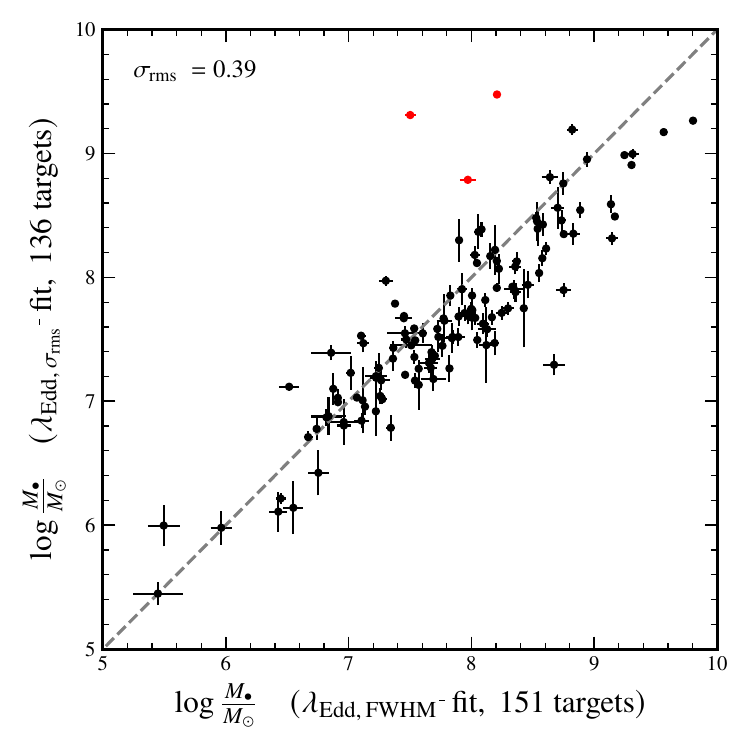}
\includegraphics[width=0.30\textwidth]{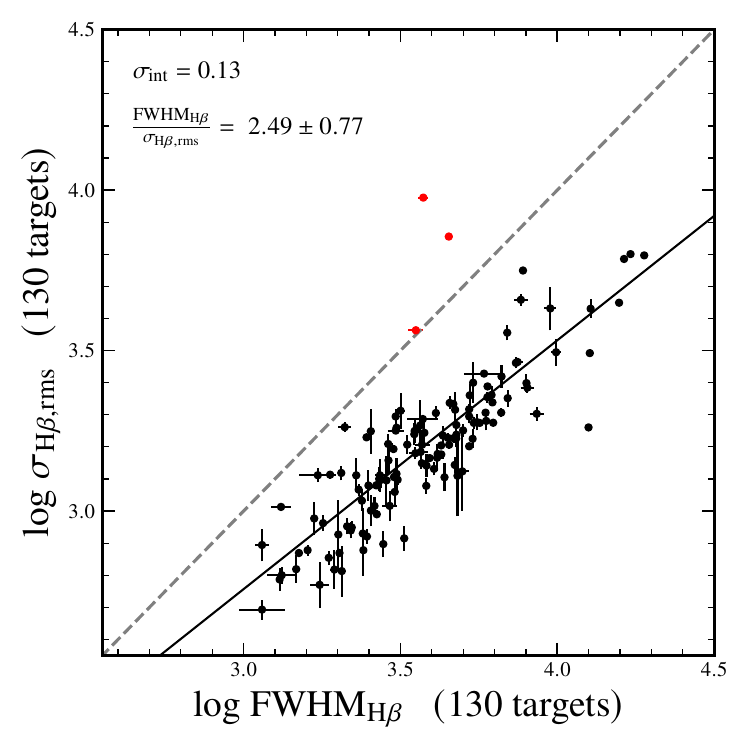}
\caption{Comparison of the expected \lag\ (left) and \mbh\ (middle), using either \vfwhm\ or \vsigrms\ as gas velocity. While the expected \lag\ shows 0.10 dex rms scatter, the mass estimate shows a larger scatter (i.e., 0.39 dex), which is mainly caused by the intrinsic scatter between \vfwhm\ and \vsigrms\ (right). We denote three targets, which are largely deviating from the one-to-one relationships with red symbols, while these targets do not show a larger offset from the fundamental plane. 
\label{fig:RL_Sigma}
}
\end{figure*}

\begin{figure*}[ht!]
\centering
\includegraphics[width=0.30\textwidth]{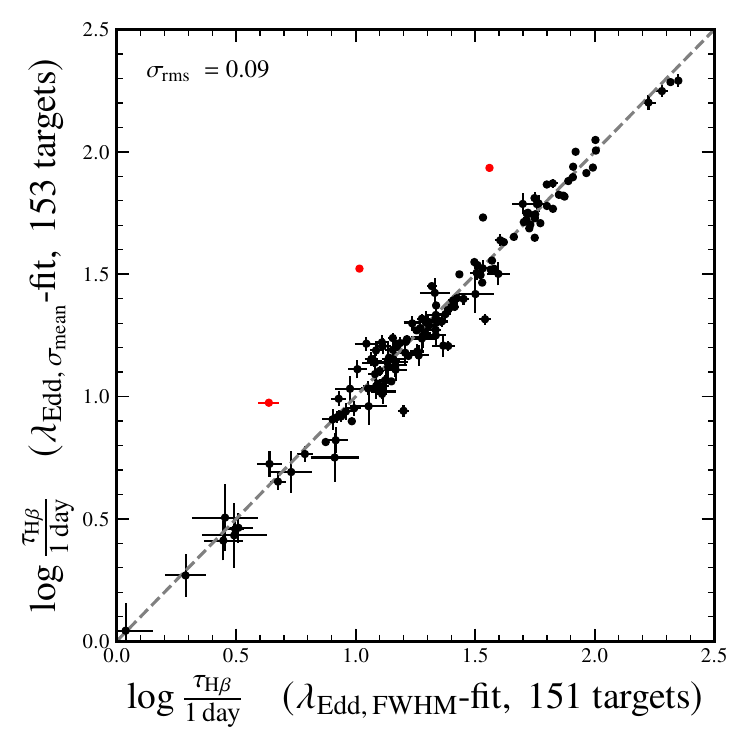}
\includegraphics[width=0.30\textwidth]{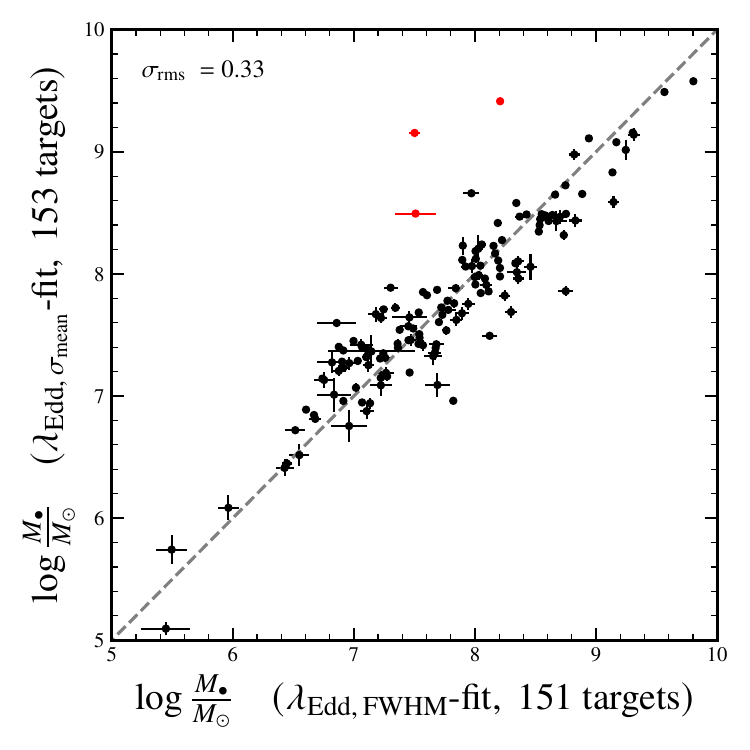}
\includegraphics[width=0.30\textwidth]{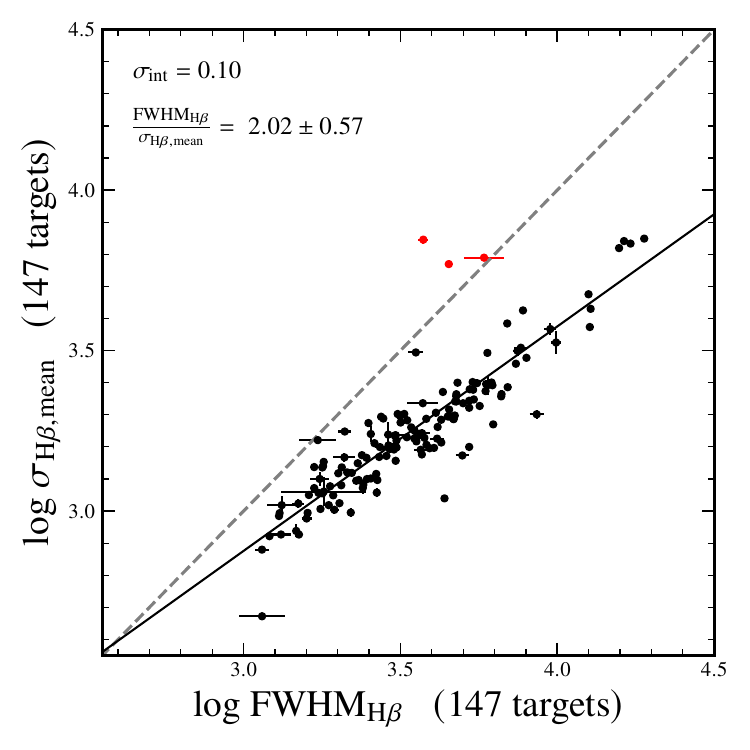}
\caption{Comparison of the expected \lag\ (left) and \mbh\ (middle), using either \vfwhm\ or \vsigmean\ as gas velocity. While the expected \lag\ shows 0.09 dex rms scatter, the mass estimate shows a larger scatter (i.e., 0.33 dex), which is mainly caused by the intrinsic scatter between \vfwhm\ and \vsigmean\ (right). We denote three targets, which are largely deviating from the one-to-one relationships with red symbols, while these targets do not show a larger offset from the fundamental plane. 
\label{fig:RL_Sigma}
}
\end{figure*}

\renewcommand{\arraystretch}{1.5}
\begin{deluxetable*}{cllcccccc}[!ht]
\tablewidth{0.99\textwidth}
\tablecolumns{8}
\tablecaption{\hb{} Single-Epoch Mass Estimators \label{table:hb_se} }
\tablehead
{
   \colhead{ID} & \colhead{$X$} & \colhead{$\Delta$V} & \colhead{sample size}  & \colhead{$p$} & \colhead{$q$} & \colhead{\intscat} & \colhead{C$_{M}$}   \\
    \colhead{(1)} &\colhead{(2)} &\colhead{(3)}&\colhead{(4)}&\colhead{(5)}&\colhead{(6)}&\colhead{(7)}&\colhead{(8)} 
}
\startdata
Case 1& $\displaystyle\eddr$    & \vfwhm & 151  & $0.48\pm 0.04$ & $2.68\pm 0.11$ & $0.21\pm0.02$ & $6.37\pm 0.22$  \\
Case 2& $\displaystyle\eddr$    & \vsigrms   & 136   & $0.46\pm 0.05$ & $2.51\pm 0.13$ & $0.21\pm0.02$ & $7.26\pm 0.21$   \\
Case 3& $\displaystyle\eddr$    & \vsigmean   & 153   & $0.47\pm 0.05$ & $2.89\pm 0.15$ & $0.21\pm0.02$ & $7.11\pm 0.14$  \\
\enddata
\tablecomments{Column (1): Case number. Column (2): The third parameter ($X$) used in Eq.~\ref{eq:relation}. Column (3): Choice of velocity. Column (4): Sample size. Columns (5): Best-fit coefficient of \lumcont\ in Eq.~11. Columns (6) : Best-fit coefficient of velocity (either \vfwhm\ or \vsig\ depending on the choice of $\Delta$V) in Eq.~11. Columns (7): Intrinsic scatter of the relation. Columns (8): Constant C$_{M}$ in Eq.~11 based on the assumed f-factor, i.e., $\log f(\vfwhm)=0.05\pm0.12$ and $\log f(\vsig)=0.65\pm0.12$ from \citet{Woo+15}. Note that the representative values are the maximum posterior estimators, and the uncertainties are the central 68\% intervals.}
\end{deluxetable*}

\begin{figure*}[ht!]
\centering
\includegraphics[width=0.80\textwidth]{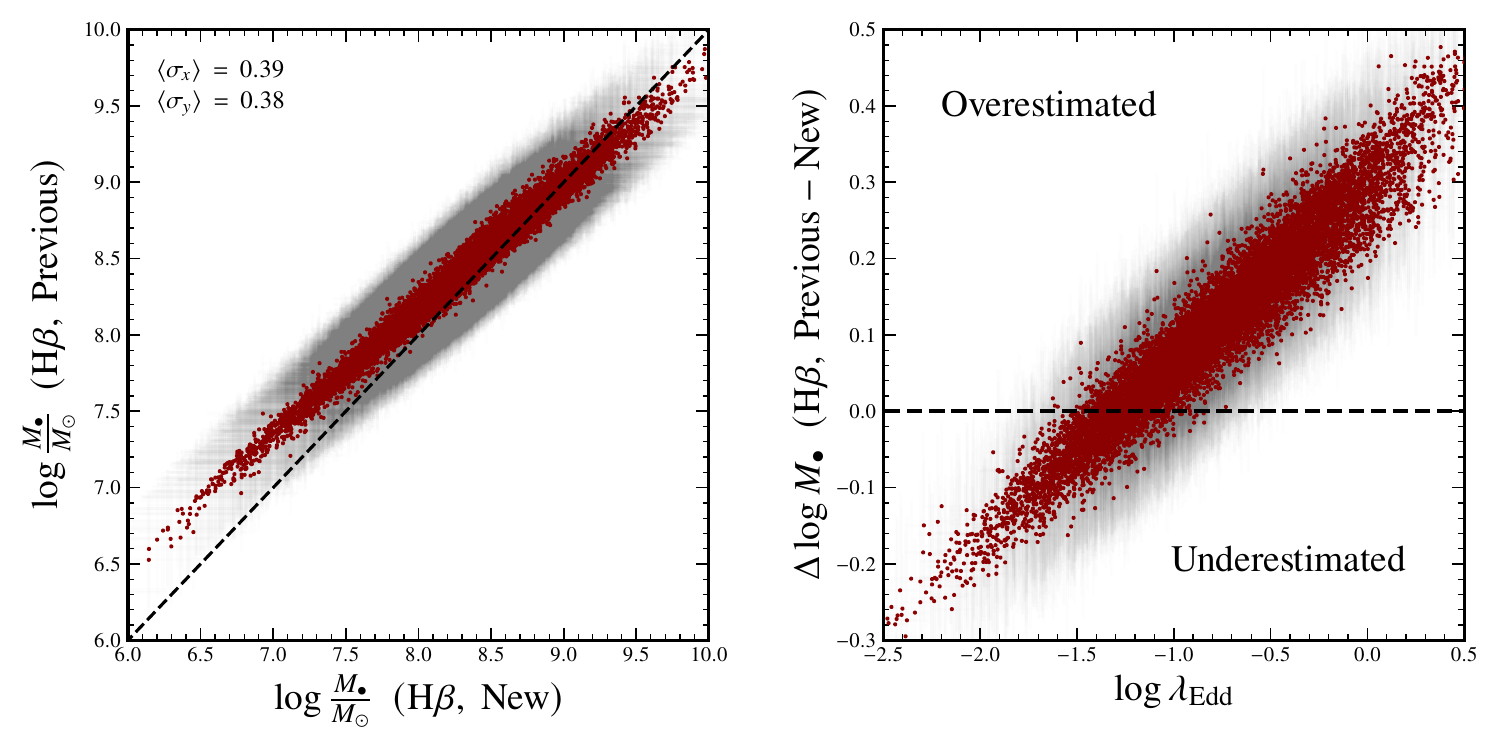}
\caption{Left: Comparison of the previous and new single-epoch mass estimates for a sample of SDSS AGNs, using \vfwhm\ and \lumcont. The uncertainties of mass estimates are calculated by adding the measurement errors to the systematic uncertainties (see Section 4.4), and the average uncertainties of the previous and new mass estimates are shown in the top-left conner. Right: A systematic trend of the overestimation of previous mass estimates as a function of Eddington ratio. For clarity, we only show the combined measurement uncertainty of \lumcont\ and \vfwhm.
\label{fig:hb_se_masscomp}}
\end{figure*}

\subsection{\mgii\ mass estimator}

To derive \mbh\ estimators using \mgii\ line velocity and UV continuum luminosity, we 
adopt the following relation from \citet{Rakshit+20}, to convert \lumuv\ to \lumcont\ and \vfwhmmg\ to \vfwhm. 
\begin{align}
\begin{aligned}
\log \lumcont  = 0.81\log \lumuv + 0.04\\
\end{aligned}
\label{eq:hb_se_vfwhm}
\end{align}
\begin{align}
\begin{aligned}
\log \vfwhm  =  1.02 \log \vfwhmmg - 0.02 \\
\end{aligned}
\label{eq:hb_se_vfwhm}
\end{align}
Then, we derive the \mgii-based \mbh\ estimator using Eq. 12 combined with Eq. 15 and 16 as
\begin{align}
\begin{aligned}
\log \mbh  = 6.35\pm0.23 +\left(0.39\pm 0.03\right)\log \lumuv  \\
+ \left(2.72 \pm 0.11\right) \log \vfwhmmg
\end{aligned}
\label{eq:mbh_mgii}
\end{align}

We  investigate the systematic effect of the new UV \mbh\ calibration using the type 1 AGN catalogue of \citet{Rakshit+20}.  
By selecting a sample of $\sim$90,000 AGNs, of which \vfwhmmg\ and \lumuv\ are available with measurement errors less than 10\%, we 
calculate \mbh\ using Eq. 17. We also estimate the mass based on the old calibration using the two-parameter size-luminosity relation of \citet{Bentz+13} combined with the conversion relations of Eq. 15 and 16. 

In Figure 9 we compare the previous mass estimates with the new ones, finding a clear systematic trend. As expected there is a clear systematic offset with Eddington ratio. In particular, the previous mass of high Eddington AGNs is overestimated by up to 0.5 dex compared to the new mass estimate. Thus, it is important to use the new calibration for properly estimating \mbh\ of high-z AGNs, for which the rest-frame UV continuum and Mg II line are available.

\begin{figure*}[ht!]
\centering
\includegraphics[width=0.80\textwidth]{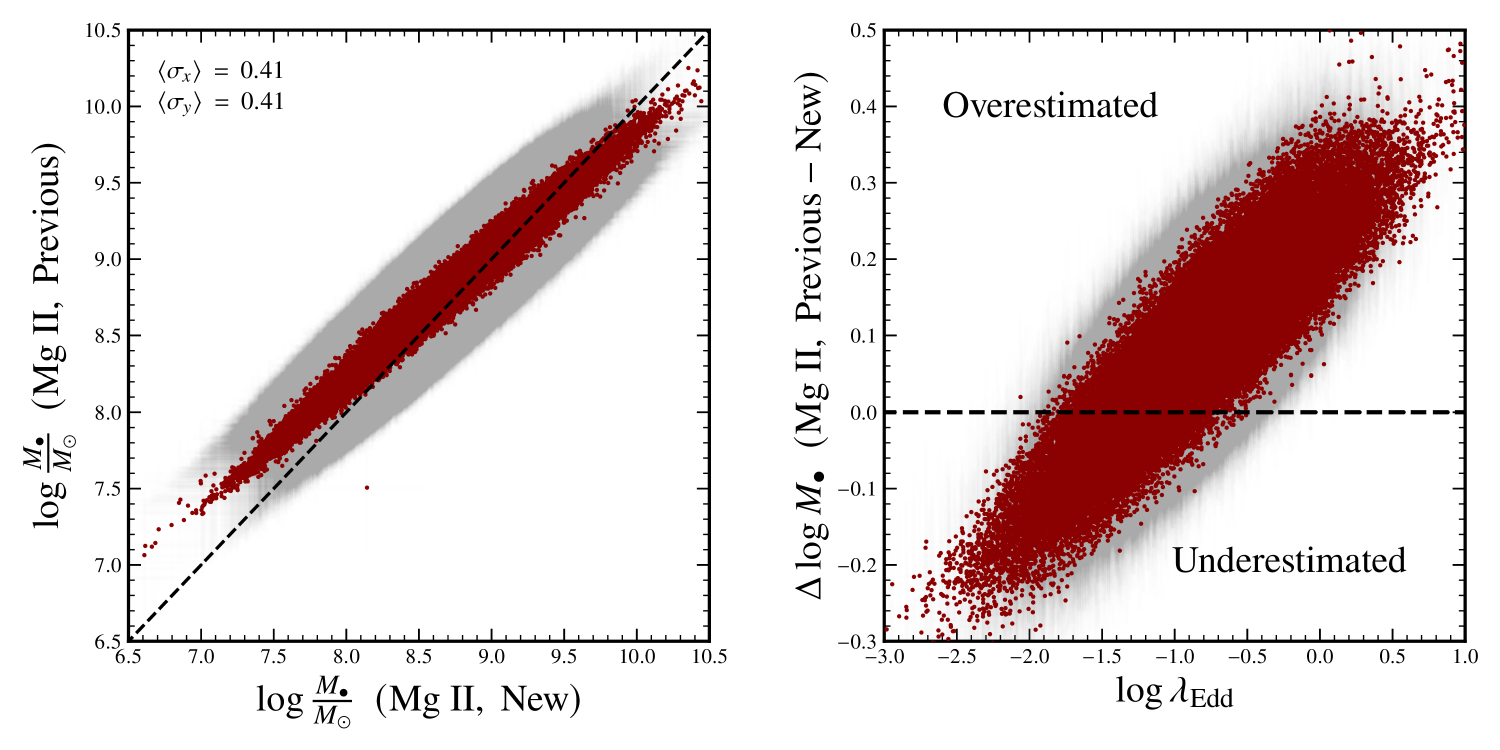}
\caption{Left: Comparison of the previous and new single-epoch mass estimates for a sample of SDSS AGNs, using \vfwhmmg\ and \lumuv. The uncertainties of mass estimates are calculated by adding the measurement errors to the systematic uncertainties (see Section 4.4), and the average uncertainties of the previous and new mass estimates are shown in the top-left conner. Right: A systematic trend of the overestimation of previous mass estimates as a function of Eddington ratio. For clarity, we only show the combined measurement uncertainty of \lumuv\ and \vfwhmmg.
}
\end{figure*}

\subsection{Correction for various f-factors}

In the previous sections, we adopt  a virial factor, $\log f=0.05\pm0.12$ for \vfwhm{} and $\log f=0.65\pm0.12$ for \vsig{} from \citep{Woo+15}. Since the Eq. 10 is derived using $\gamma$, which includes the value of $\log f$, the change of $\log f$ requires a new $\gamma$ in the three-parameter fitting analysis. In other words, if a different value of $\log f$ is used for \mbh\ determination, the $C_M$ should be corrected for since a new $\gamma$ changes the constant, C$_M$ in the mass estimators. 

We derive a correction term of C$_M$ in Eq. 11, as $\Delta$C$_M$= $\Delta \log f$,
where the virial factor difference is defined as $\Delta \log f$ = $\log f_{new} - \log f$. This correction term can be added to C$_M$ in the mass estimators (see Table 2).

\subsection{Uncertainties of single-epoch mass}

As shown in Figure~8, our new calibration substantially improves the single-epoch black hole mass estimates by 
correcting the systematic trend introduced by the Eddington ratio effect. 
In this section, we discuss the various sources of uncertainty associated with \mbh\ estimates based on the fundamental plane.
The single-epoch mass based on \hb\ can be determined from a pair of \lumcont\ and gas velocity (i.e., \vfwhm\ or \vsig) using a fundamental plane relation. By combining the measurement errors of \lumcont\ ($\delta_{\lumcont}$) and gas velocty ($\delta_{\Delta V}$), the systematic uncertainty of the predicted \lag, and the systematic uncertainty of the virial factor, we express the total uncertainty as
\begin{align}
\begin{aligned}
\delta_{\mbh} = \sqrt{ \alpha^2 \delta_{\lumcont}^2 + \beta^2 \delta_{\Delta V}^2 +  \delta_{FP}^2 +  \delta_{f}^2 }
\end{aligned}
\end{align}
where $\alpha$ and $\beta$ come from Eq. 3.
First, the measurement errors ($\delta_{\lumcont}$ and $\delta_{\Delta V}$) are often negligible for high quality data. However, these uncertainties could be significant when the signal-to-noise of the single-epoch spectra is relatively low. Thus, the measurement uncertainties should be added to other systematic uncertainties. 
Second, the predicted \lag\ has a systematic uncertainty ($\delta_{FP}$), for which we adopt an intrinsic scatter of 0.21 dex from the fundamental plane relations in Table 1. Third, we account for the uncertainty caused by the average virial factor ($\delta_f$), since individual AGNs may have different virial factors. Using a sample 35 AGNs with velocity-resolved lag measurements, for example, Wang et al. (in preparation) reported an rms scatter of $\sim0.3$ dex between RM-based \mbh\ and the \mbh\ determined from dynamical modeling \citep[see also][]{Pancoast+14, Williams+18, Villafana+23}. We adopt this scatter as the systematic uncertainty associated with the average virial factor. Note that various previous studies adopted an uncertainty of $\sim0.2-0.3$ dex for the virial factor based on the scatter of the \mbh-stellar velocity dispersion relation of the reverberation sample \citep[e.g.,][]{Park+12, Woo+13}.

Excluding measurement uncertainties, we obtain the combined uncertainty of the \hb-based single-epoch mass with \vfwhm\ as 0.37 dex, by adding $\delta_{FP}$ (0.21 dex) and $\delta_{f}$ (0.3 dex) in quadrature. The measurement uncertainties ($\delta_{\lumcont}$ and $\delta_{\vfwhm}$) should be added to the combined error in quadrature for individual AGNs. For example, if the combined measurement uncertainty is 0.1 dex, the total mass uncertainty is 0.38 dex. 

For comparison, the RM-based mass uncertainty is somewhat smaller due to the direct measurement of \lag. If we assume a 0.1 dex measurement error of \lag\ and the combined measurement uncertainty of $\delta_{\lumcont}$ and $\delta_{\vfwhm}$ as 0.1 dex, the total RM-based mass uncertainty is 0.33 dex by adding the measurement errors to $\delta_f$ in quadrature. Note that the systematic uncertainty of the virial factor applies to both RM-based and single-epoch masses. 

If we take more conservative approach, we need to account for the systematic difference due to the choice of gas velocity since both single-epoch or RM-based masses can differ by 0.3-0.4 dex depending on the choice of either \vfwhm\ or \vsig\ in the virial equation, as shown in Figure 6 and 7. Therefore, this uncertainty should be included for a conservative estimate of the mass uncertainty. 

For the \mgii-based single-epoch mass estimates, we additionally consider the uncertainty in the conversion from \lumuv\ to \lumcont\ (Eq. 15) and from \vfwhmmg\ to \vfwhm\ (Eq. 16). Using a subsample of AGNs presented in Section 4.2, for which both \hb\ and \mgii\ FWHM measurements and both \lumcont\ and \lumuv\ are available, we obtain intrinsic scatters of $\sim0.08$ dex and $\sim0.11$ dex, respectively, for the luminosity and line FWHM comparisons. We adopt these scatters as systematic uncertainties of the two conversions. 
Excluding the measurement uncertainties, the combined error of the \mgii-based mass is 0.39 dex by combining the two conversion uncertainties to the uncertainty from Eq. 12 (i.e, 0.37 dex). The total mass uncertainty would be obtained by combining 0.39 dex and the measurement uncertainties of \lumuv\ and \vfwhmmg.

\section{Discussion}

As reported by various recent studies \citep{Du+18, Woo+24}, high Eddington AGNs show a clear offset from the \hb\ BLR size-luminosity relation. In this work, we empirically fit the three-parameter correlation by introducing either Eddington ratio or gas velocity as the third parameter. Based on these experiments, we find the fundamental plane defined in the 3-D space of BLR size, continuum luminosity, and gas velocity. The tilt of the fundamental plane is set to the Eddington ratio direction, i.e., the Eddington ratio effect is minimized since most AGNs lie on the plane while the Eddington ratio increases or decrease along the tilt of the plane. The fundamental plane indicates that the size of BLR is not simply determined by photoionization but also connected with kinematics although it is yet to be clear what causes the reduced BLR size for higher Eddington AGN compared to lower Eddington AGN at a fixed luminosity \citep[see discussion by][]{Woo+24}.

Nevertheless, we empirically define tight correlations with three-parameters, enabling for correcting for the Eddington ratio effect. The expected BLR size (or \lag) is consistent regardless of the calibrations and the \mbh\ estimate is also self-consistent. 

On the other hand, the new mass estimates show a clear systematic offset from the previous mass estimates that based on the two-parameter size-luminosity relation. In particular, this trend is substantially large for high Eddington AGNs, indicating the previous mass estimates could be overestimated by a factor of up to 3. For high-z AGNs, the rest-frame UV lines are typically observed and used for single-epoch \mbh\ estimation. We showed that \mgii-based mass also suffers the same systematic trend that the mass of high Eddington AGN can be overestimated by up to a factor of 3. Note that the systematic offset depends on the details of the \mgii-based \mbh\ calibration since the measured \lumuv\ has to be converted to \lumcont\ and the measured velocity of \mgii\ line (\vfwhmmg) has to be converted to \hb\ velocity (\vfwhm). Note that there are various calibration between UV and optical luminosities and between \vfwhmmg\ and \vfwhm, which often show inconsistent results \citep[i.e., different slopes, see][]{Wang+09, Bahk+19, Wang+19, Le+20}. Therefore, \mgii-based \mbh\ is in general more uncertain due to these additional calibration.

\subsection{\Rfe\ as the third parameter}

\begin{figure}[ht!]
\centering
\includegraphics[width=0.48\textwidth]{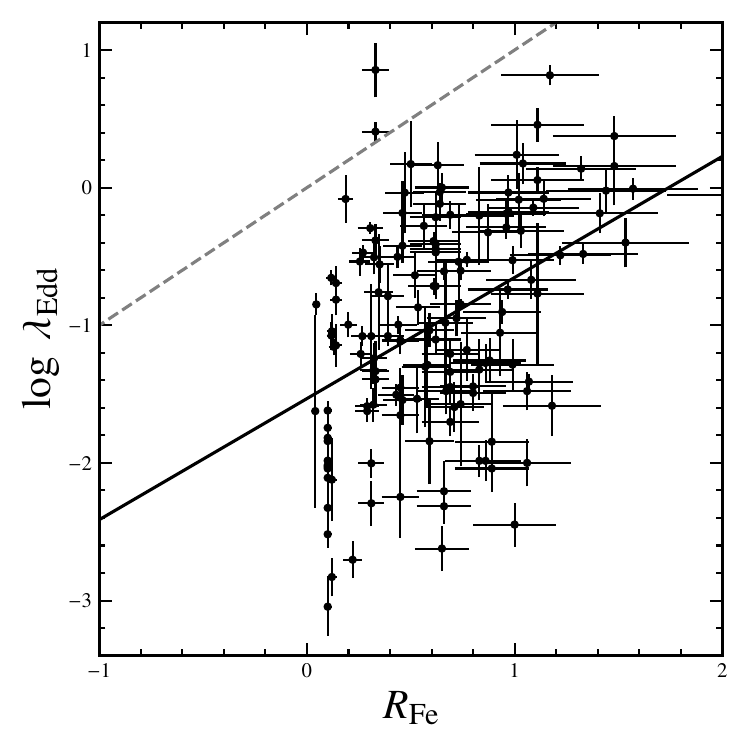}
\caption{Comparison of \Rfe\ with the Eddington ratio. While \Rfe\ broadly correlates with the Eddington ratio, there is substantially large scatter along the best-fit (solid line), indicating \Rfe\ is not a good proxy for the Eddington ratio. The one-to-one relationship is denoted with a dashed line.
\label{fig:Rfe}}
\end{figure}
 
 \begin{figure}[ht!]
\centering
\includegraphics[width=0.45\textwidth]{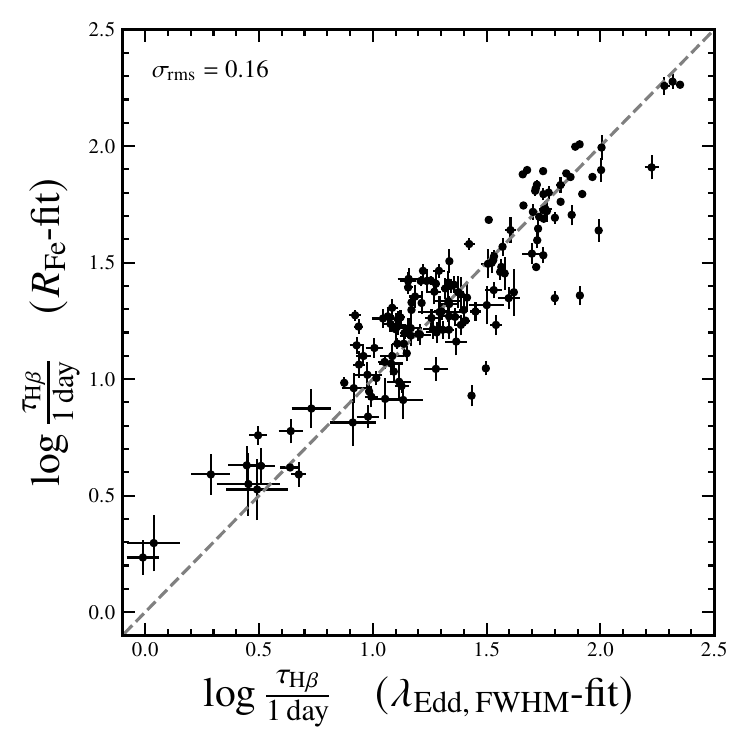}
\includegraphics[width=0.45\textwidth]{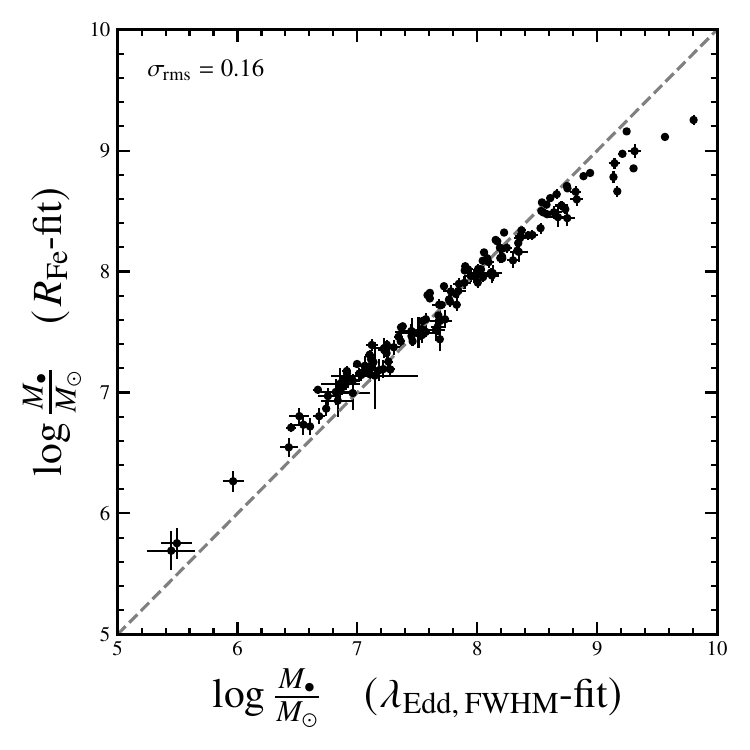}
\caption{Comparison of the predicted \lag\ (top) and \mbh\ (bottom) based on the three-parameter fit with \Rfe\ compared to the predictions based on the direct measurements of Eddington ratio.The three-parameter fitting with \Rfe\ predicts systematically different values, leading to a large uncertainty of \lag\ and \mbh.
}
\end{figure}

As discussed in Section~1, previous studies have used various observable indicators of the Eddington ratio to correct for the offset of high-Eddington AGNs from the canonical BLR size–luminosity relation. For example, \citet{Du&Wang19} adopted an empirical approach, exploring multiple observables and reported that R$_{\rm Fe}$ provides the strongest three-parameter correlation with $R_{\rm BLR}$ and $L_{5100}$. Given the similarity between their approach and ours, we examine whether R$_{\rm Fe}$ can serve as an alternative third parameter and compare the resulting fits with those based on the Eddington ratio.

While R$_{\rm Fe}$ has been widely used as a proxy for the Eddington ratio, several limitations are well known. 
First, the integrated Fe\,{\sc ii} flux between 4434\,\AA\ and 4684\,\AA\ is subject to uncertainties due to template mismatch, and individual Fe\,{\sc ii} components may vary on different timescales \citep{Kovasevic10}. 
Although Fe\,{\sc ii} emission is typically measured with available templates, systematic differences arise depending on the adopted template (e.g., \citealt{Park22}). 
Second, the narrow H$\beta$ flux is particularly uncertain in narrow-line Seyfert~1 galaxies, since the broad H$\beta$ component is relatively narrow (FWHM $<2000$ km\,s$^{-1}$), making it difficult to deblend the narrow component. 
Although the flux ratio of narrow H$\beta$ to [O\,{\sc iii}]\,5007\,\AA\ ranges from 6--16\% \citep{Marziani03}, some studies \citep[e.g.,][]{Du&Wang19} have assumed a fixed ratio, introducing additional systematics in R$_{\rm Fe}$.

Despite these limitations, we tested R$_{\rm Fe}$ as a potential third parameter by adopting measurements for a subsample of 148 AGNs with available FWHM$_{\rm H\beta}$ from our own measurements or the literature. 
Because individual measurement uncertainties are not available for most sources (see Du \& Wang 2019) provided that the adopted 20\% error in R$_{\rm Fe}$ is realistic. 

In Figure~10 we compare R$_{\rm Fe}$ and the Eddington ratio, showing that the correlation is weak and exhibits a large scatter of 0.69\,dex. 
For a given R$_{\rm Fe}$, the Eddington ratio spans nearly three orders of magnitude, indicating that the limited dynamic range of R$_{\rm Fe}$ contributes to its poor performance as a reliable proxy. 
  
Using R$_{\rm Fe}$ as the third parameter, and adopting FWHM$_{\rm H\beta}$ as the velocity measure in Eq.~1 for computing $M_\bullet$ and $\lambda_{\rm Edd}$, we performed a three-parameter fit adopting the same orthogonal procedure used in Section 3.1 and obtained
\begin{align}
\begin{aligned}
\log \lag= \left(0.46\pm 0.02\right)\log \lumcont \\
- \left(0.23\pm 0.06\right) R_{\rm Fe} + \left(1.55\pm 0.04\right)
\label{eq:bestfit_Rfe}
\end{aligned}
\end{align}
with an intrinsic scatter of $\intscat=0.19\pm0.02\,\mathrm{dex}$. 
This result is similar to that of \citet{Du&Wang19} and the intrinsic scatter is comparable to the case of the \eddr-fit in Eq.~ 4 if we assume the adopted 20\% error of \Rfe\ is realistic. 
We note that \citet{Du&Wang19} used \Rfe\ in linear scale in their fitting, and we tested both \Rfe\ and $\log R_{\rm Fe}$, finding comparable results. 

Despite the apparently comparable scatter, the predicted $\tau_{\rm H\beta}$ values from the R$_{\rm Fe}$ fit differ systematically from those based on the $\lambda_{\rm Edd}$-fit.
As shown in Figure~11, the R$_{\rm Fe}$-based correlation tends to overestimate or underestimate $\tau_{\rm H\beta}$, leading to potential biases in the inferred $M_\bullet$. 
These discrepancies demonstrate that the R$_{\rm Fe}$ fit provides no better prediction compared to the case of the $\lambda_{\rm Edd}$ fit.

\subsection{Implications to BH growth}

The new mass calibration indicates substantially lower \mbh\ for high Eddington AGNs. This may lower the tension of the fast growth of high-z AGNs from their seeds. In Figure 10 we demonstrate this effect by calculating \mgii-based mass of high-z AGN at $6< z < 8$ adopted from \citet{Inayoshi20}, who collected high-z AGN masses. Note that we only select high-z AGNs, for which \mgii\ line based mass can be calculated. Compared to the original calculation by the author, we show substantially lower masses, which is consistent with the Eddington-limited growth of 100 \msun\ BHs. As a reference, we also show the growth line for a BH seed of 10$^4$ \msun\ with a 50\% duty cycle. These demonstrations indicate that reliable \mbh\ mass estimates are required to properly understand the nature of BH seeds and their growth. 

\begin{figure}[ht!]
\centering
\includegraphics[width=0.48\textwidth]{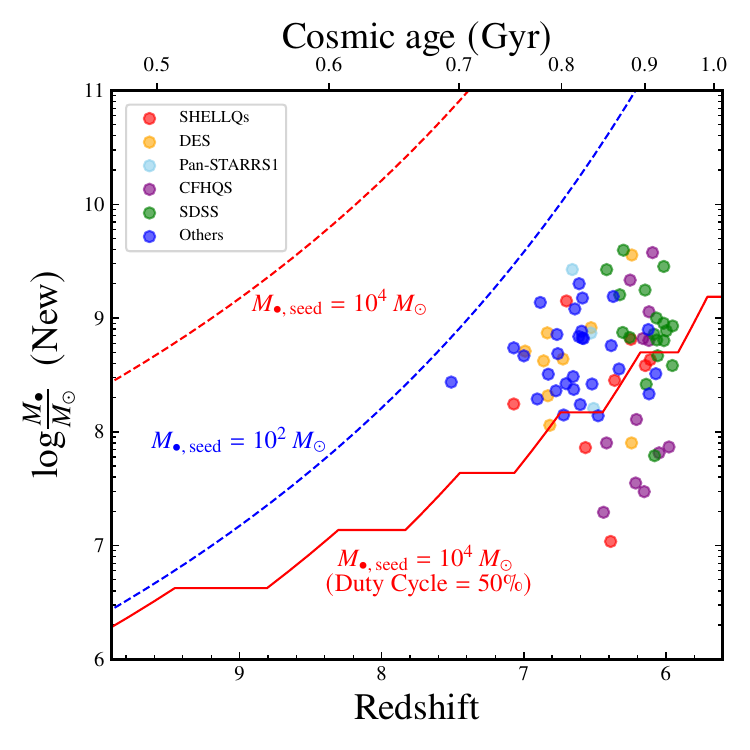}
\caption{Models of the BH seed growth compared to the high-z AGNs. New \mbh\ is calculated for the AGNs from \citet{Inayoshi20}, who collected various samples \citep[see reference by][]{Inayoshi20}. Note that we only present the \mgii-based \mbh. The Eddington-limited growth of a BH seed with 10$^4$ \msun\ (red dashed line) or 10$^2$ \msun\ (blue dashed line) is compared to the growth of duty cycle 50\% of  a 10$^4$ \msun\ BH (red solid line). 
}
\end{figure}

\section{Summary and conclusions}

To address the systematic offset of high Eddington AGNs from the BLR size-luminosity relation, we investigate the dependence of the BLR size with other AGN parameters, using a robust sample of 157 AGNs with reliable \hb\ lag measurements from \citet{Wang&Woo24}. Our main conclusions are as follows:

$\bullet$ By introducing the Eddington ratio as the third parameter, we define a fundamental plane in the three-dimensional space of \lag, \lumcont, and \vfwhm\ (or \vsig). At fixed luminosity, the plane reveals a positive correlation between the BLR size and gas velocity, indicating that its tilt is defined along the Eddington ratio, effectively accounting for the Eddington ratio effect.

~

$\bullet$ Using the fundamental plane, the BLR size can be predicted from a pair of optical luminosity and gas velocity measurements within an intrinsic scatter of $\sim0.21$ dex. The predicted \lag\ values are consistent within $\sim0.1$ dex, regardless of the choice of the calibrations performed with various gas velocities (i.e., \vfwhm, \vsigrms, or \vsigmean).

~

$\bullet$ We provide new single-epoch \mbh\ estimators based on the fundamental plane. The uncertainty of single-epoch mass arising from \lag\ prediction is 0.21 dex, while single-epoch masses are consistent within 0.1 dex regardless of the choice of the calibrations, highlighting the robustness of this method. However, the difference in velocity measurements (i.e., 0.1-0.13 dex between \vfwhm\  vs. \vsigrms\ (or \vsigmean) introduces an additional $\sim0.3$ dex uncertainty, which is applied to both reverberation-based and single-epoch masses. 

~

$\bullet$ The new mass estimators substantially reduce \mbh\ of high Eddington AGNs by up to a factor of 3, 
demonstrating the importance of using the three-parameter relation.
The calibration significantly lowers the mass of high-z AGNs, alleviating the tension in BH seed growth scenarios. 

~

The new empirically defined fundamental plane underscores the close connection between BLR gas kinematics and the photoionization. Future theoretical work is required to understand the physical origin of this relation and the reduced BLR size in high Eddington AGNs.

%%%%%%%%%%%%%%%%%%
% Acknowledgment %
%%%%%%%%%%%%%%%%%%

\begin{acknowledgments}
We thank the anonymous referee for valuable suggestions. 
This work has been supported by the Basic Science Research Program through the National Research Foundation of Korean Government (2021R1A2C3008486).
\end{acknowledgments}

\appendix

\renewcommand{\theequation}{A\arabic{equation}}

\section{Correlation dependency}

To properly calculate the error of each parameter in the three-parameter fitting, we propagate the measurement uncertainties and consider whether measurements are correlated or not. 
For example, \eddr{} is not independently measured but computed from the  5100\angstrom{} luminosity and the rest-frame \hb{} time lag, the measurement uncertainties of these three variables should be correlated. Ignoring the redshift uncertainties, the uncertainty correlations can be expressed as
\begin{align}
\corr{\uncert{\lag}}{\uncert{\eddr}}&=-\frac{\uncert{\lag}}{\uncert{\eddr}} \label{eq:ucorr_te} \\
\corr{\uncert{\lumcont}}{\uncert{\eddr}}&=\frac{\uncert{\lumcont}}{\uncert{\eddr}} \label{eq:ucorr_le} \\
\corr{\uncert{\lag}}{\uncert{\lumcont}}&=0 \label{eq:ucorr_tl}
\end{align}
where $\uncert{x}$ denotes the measurement uncertainty of $\log x$ and $\corr{x}{y}$ denotes the Pearson correlation coefficient. Note that line widths were measured independently from \hb{} lag or \lumcont{}, thus having no correlation between them.
In principle, \lumcont{} and rest-frame \lag{} are also correlated, given that they are computed from the flux and observed-frame \lag{} with redshift. However, we note that the contribution of redshift uncertainty to the correlation of uncertainties is insignificant since the redshift uncertainties are negligible compared to the uncertainties of flux and observed-frame lag. Thus, we ignore the correlations arising from the redshift uncertainties.

We experiment with the \eddr-fit assuming that Eddington ratio is measured independently from the \hb{} time lag or gas velocity (line width). We find that the fitting results are consistent with the correlated uncertainties. While the intrinsic scatter is marginally reduced since independent uncertainties overestimate the measurement uncertainties, we find consistent results compared to the case with correctly modeling uncertainty correlations. 

\section{three-parameter fitting comparison}

As described in \S~3, the dependency among the three parameters, i.e., \Hb\ time delay, luminosity, and Eddington ratio (or velocity) is not clearly defined, and we presented the best-fit fundamental plane results based on the rotation-invariant fitting scheme. In this section, for completeness, we present the three-parameter correlation using conditional regression.
Conditional regression minimizes the residuals along the axis of the dependent variable. This approach is often adopted when the goal of the regression is to predict one parameter from the other, even if the physical dependency is not fully established. By construction, the scatter can be minimized along the chosen axis; however, the inferred correlation may be attenuated relative to a symmetric regression. 

Nevertheless, we test the three-parameter correlation by alternatively assuming each parameter to be dependent on the other two and minimizing the residuals along the axis of the assumed dependent variable. In practice, we focus on the correlation among \Hb\ time delay (\lag), optical luminosity (\lumcont), and Eddington ratio (\eddr), adopting the FWHM of \Hb\ as the velocity indicator. The three-parameter relation is expressed as
\begin{align}
\log Y = \alpha \log X_1 + \beta \log X_2 + \gamma ,
\label{eq:relation}
\end{align}
where Y is the dependent variable and X$_1$ and X$_2$ are independent variables. 

For the conditional regression, we first follow the scheme presented by \citet{Du&Wang19}, implementing a Levenberg-Marquardt optimization algorithm and using bootstrap resampling to estimate parameter uncertainties. The minimization is performed along the axis of the assumed dependent variable (i.e., the Y-axis). Because the regression is performed in a single direction, the intrinsic scatter perpendicular to the plane is not explicitly constrained. We therefore report the rms scatter along the $Y$-axis. For comparison, we also compute the rms scatter orthogonal to the best-fit plane.

Second, we construct an initial implementation of a 3-D extension of \textsc{linmix\_err} for three-parameter regression, since the available \textsc{Python} code is limited to two-parameter fitting. We note that this implementation represents a preliminary extension and has not yet undergone full validation. A direct generalization of \textsc{linmix\_err} to three dimensions is technically non-trivial. In two-dimensional regression, \textsc{linmix\_err} employs a Gaussian mixture model prior on the latent variable, which enables analytic marginalization and leads to an efficient Gibbs sampling scheme. In a three-dimensional regression, however, the latent predictors form a multidimensional vector rather than a scalar, requiring a multivariate Gaussian mixture prior with full covariance matrices for each mixture component. The number of free parameters therefore increases substantially, and sampling these covariance matrices becomes considerably more complex and potentially unstable. In our simplified initial version, we adopt several assumptions regarding the prior structure, which requires further validation. Accordingly, the results based on our modified version of the \textsc{linmix\_err} are presented for comparison purpose only (see Table 3). Note that we obtain similar fitting results compared to those based on the \citet{Du&Wang19} method.

In Figure 13, we present the best-fit relations based on the conditional regression using the scheme of \citet{Du&Wang19}. When \lag\ is assumed to be the dependent variable (Figure 13, left), we obtain the tightest correlation, comparable to that derived from the rotation-invariant fit using the \textsc{Hyper-Fit} code (see Figure 2, left). In contrast, when \lumcont\ or \eddr\ is assumed to be the dependent variable (Figure 13, middle and right), the rms scatter increases substantially. 

We show the geometrical structure of  the three parameters in Figure 14. When \lag\ is treated as the dependent parameter, a well-defined fundamental plane emerges (Figure 14, left), which is broadly consistent with the plane defined from our orthogonal regression (see Figure 5). Although these results may imply that \lag\ is dependent on \lumcont\ and velocity than vice versa, the strong variation in the best-fit relations depending on the assumed dependent variable indicates that the inferred  correlation is sensitive to the regression direction. This sensitivity supports the use of a rotation-invariant regression, which better constrains the geometric structure of the three-parameter relation. 

Numerous studies have employed the two-parameter BLR size–luminosity relation for predictive purposes \citep[e.g.,][]{Vestergaard02, WooUrry02, Bentz+13}. For example, the measured \lag\ is commonly used to infer the BLR size and black hole mass, as discussed in Section~4. Conversely, the continuum luminosity can be predicted from the BLR size–luminosity relation and has been proposed as a distance indicator for cosmological applications \citep{Watson+11, Czerny+13, MartinezAldama19, Yu+23}. 
The fundamental plane relation provides an improved predictive framework compared to the traditional two-parameter BLR size–luminosity relation, 
as it incorporates an additional physical parameter and reduces the intrinsic scatter. In principle, the fundamental plane enables the prediction of one parameter 
(e.g., luminosity, time delay, or velocity) from the other two, thereby improving applications such as black hole mass estimation 
and cosmological distance measurements.
However, the inferred fundamental plane depends on the adopted regression scheme. When a specific dependent variable is assumed in conditional regression, 
the resulting correlation varies significantly with the chosen direction, indicating that the fitted plane is sensitive to this assumption. 
To avoid such directional bias, the rotation-invariant fundamental plane obtained from symmetric regression provides a more robust tool for predicting time delay, continuum luminosity, or velocity when the dependency among the three parameters is not clearly defined.

\renewcommand{\arraystretch}{1.5}
\begin{deluxetable*}{cccccccccccc}[!ht]
\tablewidth{0.99\textwidth}
\tablecolumns{12}
\tablecaption{Three-parameter fitting results for various regression methods and dependent variable assumptions \label{table:bestfit}}
\tablehead
{
    \colhead{method}&fit direction&\colhead{$Y$}&\colhead{$X_1$}&\colhead{$X_2$}&\colhead{$\alpha$}&\colhead{$\beta$}&\colhead{$\gamma$}&\colhead{\ensuremath{\sigma_\mathrm{rms,y}}}&\colhead{\ensuremath{\sigma_\mathrm{rms,orth}}}&\colhead{\ensuremath{\sigma_\mathrm{int, y}}}&\colhead{\ensuremath{\sigma_\mathrm{int,orth}}}
\\
    % \colhead{}&\colhead{}&\colhead{}&\colhead{}&\colhead{}&\colhead{}&\colhead{}
    % \\
    \colhead{(1)}&\colhead{(2)}&\colhead{(3)} &\colhead{(4)} &\colhead{(5)}&\colhead{(6)}&\colhead{(7)}&\colhead{(8)}&\colhead{(9)}&\colhead{(10)}&\colhead{(11)}&\colhead{(12)}
}
\startdata
DW19 & Y-axis & \lag & \lumcont & \eddr &  $0.46\pm 0.04$ & $ 0.56\pm 0.10$ & $1.16\pm 0.07$ & $0.26$ & $0.22$ & -- & -- \\
DW19 & Y-axis & \lumcont & \lag     & \eddr      &  $0.80\pm 0.12$ & $1.07\pm0.25$ & $0.62\pm 0.27$ & $0.44$ & $0.30$ & -- & -- \\
DW19 & Y-axis & \eddr & \lag     & \lumcont &  $0.69\pm 0.89$ & $ 7.69\pm 3.73$ & $-2.77\pm 2.47$ & $1.96$ & $0.25$ & -- & -- \\
linmix\_3D & Y-axis & \lag & \lumcont & \eddr &  $0.45\pm 0.04$ & $ 0.50\pm 0.09$ & $1.14\pm 0.07$ & -- & $0.21$ & $0.20\pm0.01$ & -- \\
linmix\_3D & Y-axis & \lumcont & \lag     & \eddr &  $0.62\pm 0.09$ & $ 1.14\pm 0.15$ & $0.76\pm 0.15$ & -- & $0.23$ & $0.35\pm0.03$ & --\\
linmix\_3D & Y-axis & \eddr & \lag     & \lumcont &  $0.61\pm 0.37$ & $ 6.45\pm 5.18$ & $-2.13\pm 3.02$ & -- & $0.25$ & $1.58\pm0.97$ & --\\
\hline
Hyper-Fit & orthogonal &\lag & \lumcont & \eddr & $0.48\pm 0.04$ & $ 0.68\pm 0.11$ & $1.03\pm 0.08$ & -- & $0.21$ & -- & $0.21\pm0.02$ \\
\enddata
\tablecomments{Column (1): The fitting method using the scheme of Du \& Wang 19 (DW19) or our modified linmix\_err (linmix\_3D). Column (2): The direction of the fitting. Column (3): The dependent parameter. Column (4): The first independent parameter. Column (5): The second independent parameter. Column (6): Best-fit coefficient of \lumcont\ in Eq.~A4. Column (7) : Best-fit coefficient of velocity (i.e., \vfwhm). Column (8): Best-fit coefficient in Eq.~\ref{eq:relation}. Column (9): rms scatter in the Y-axis. Column (10): rms scatter perpendicular to the hyperplane. Column (11): Intrinsic scatter in the Y-axis. Column (12): Intrinsic scatter perpendicular to the hyperplane. Note that the representative values are the maximum posterior estimators, and the uncertainties are the central 68\% intervals. $\log f(\vfwhm)=0.05\pm0.12$ were adopted from \citet{Woo+15} to obtain relevant fits. All conditional fitting results are based on the sample of 151 AGNs with \vfwhm. In the last raw, we repeated the orthogonal fitting result of the Case 1 in Table 3 for comparison. }
\end{deluxetable*}

\begin{figure*}[ht!]
\centering
\includegraphics[width=0.31\textwidth]{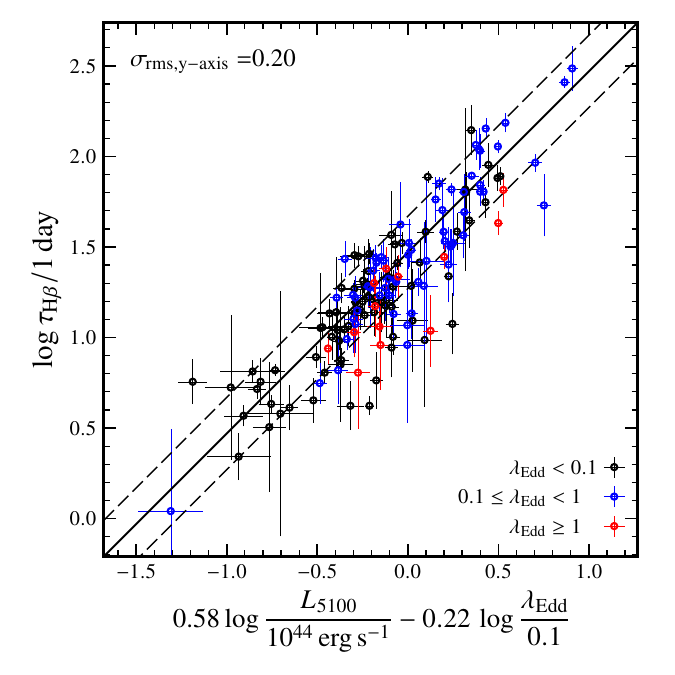}
\includegraphics[width=0.31\textwidth]{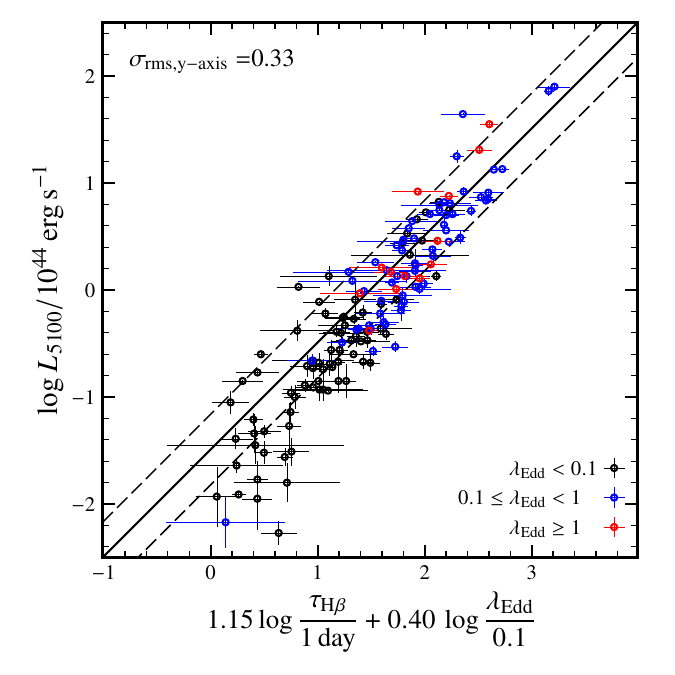}
\includegraphics[width=0.31\textwidth]{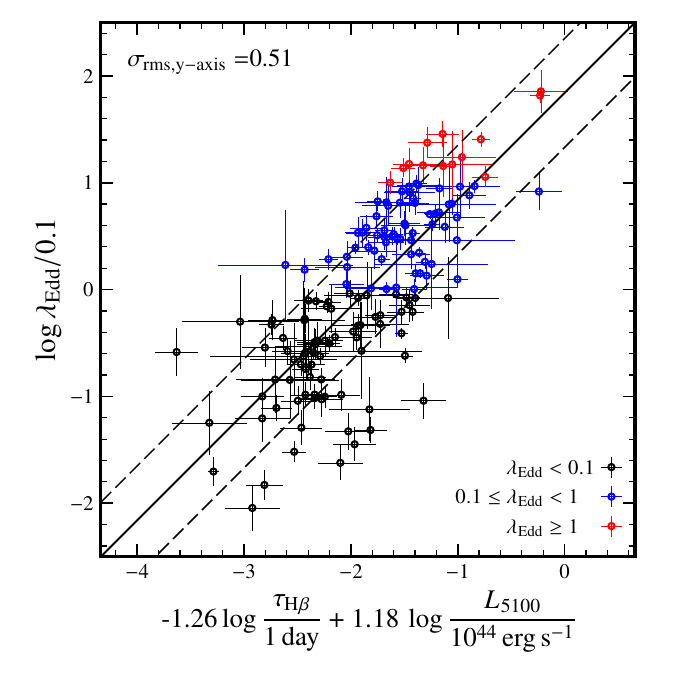}
\caption{The three-parameter fitting results obtained using the method of \citet{Du&Wang19}, assuming the variable along the Y-axis as the dependent parameter.
For comparison, see Figure 2 (left), which presents the rotation-invariant fit based on \textsc{Hyper-fit}. }
\end{figure*}

\begin{figure*}[ht!]
\centering
\includegraphics[width=0.32\textwidth]{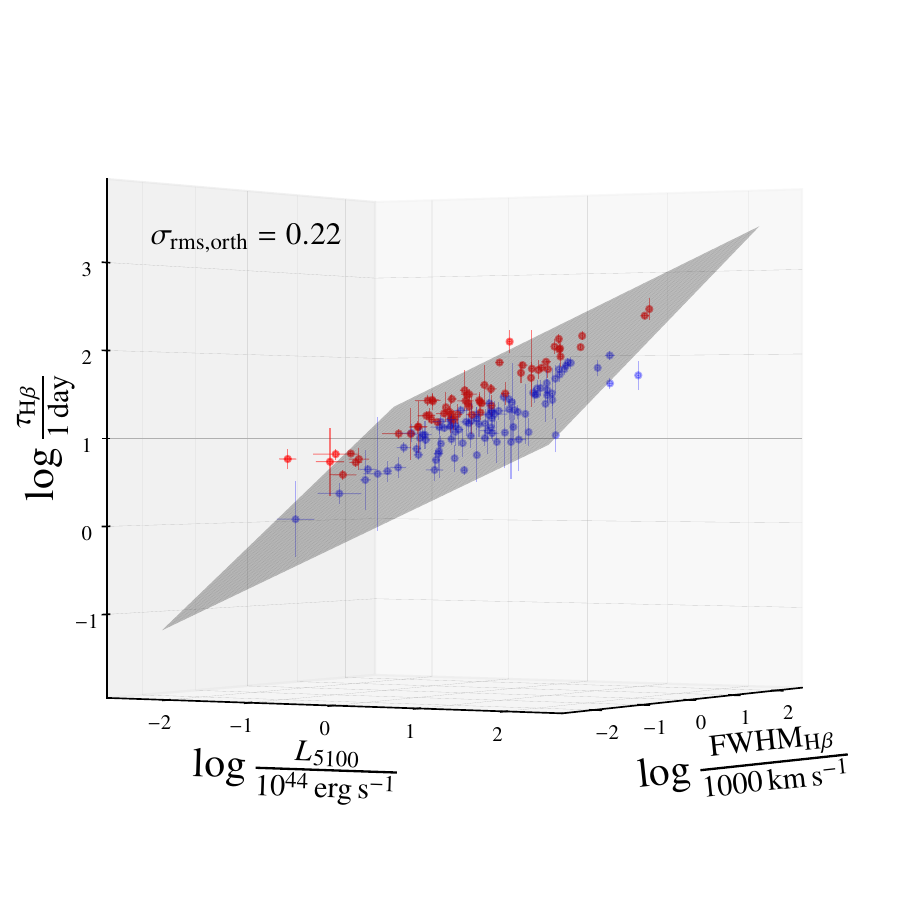}
\includegraphics[width=0.32\textwidth]{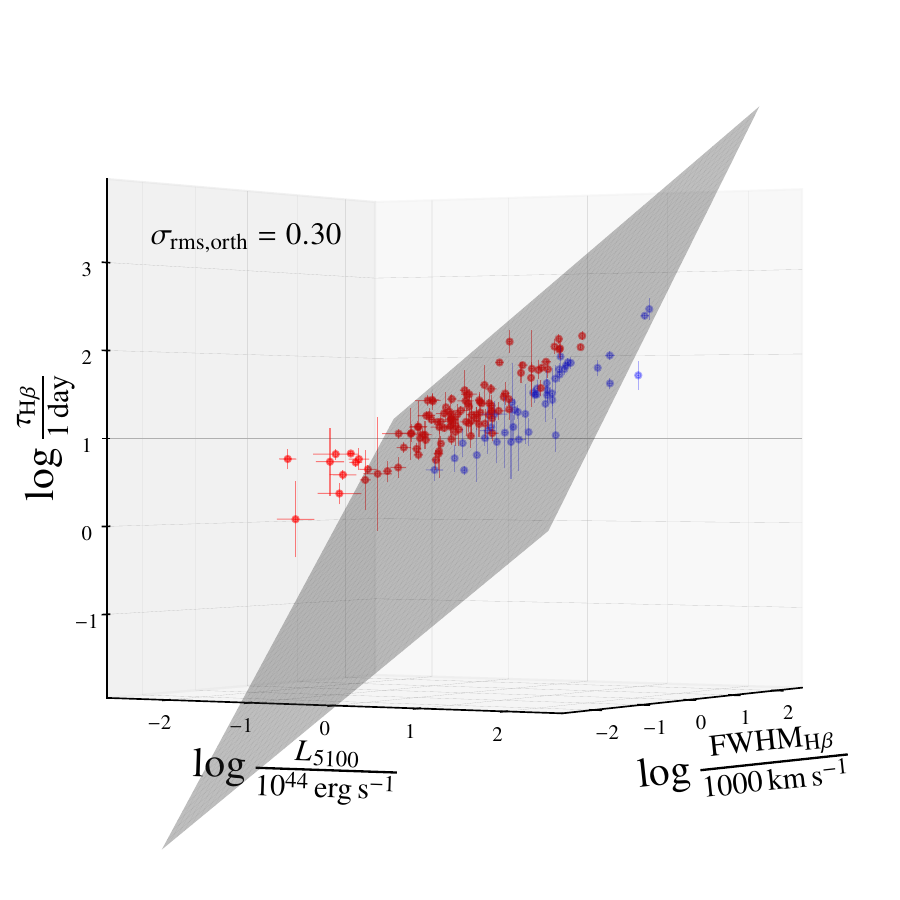}
\includegraphics[width=0.32\textwidth]{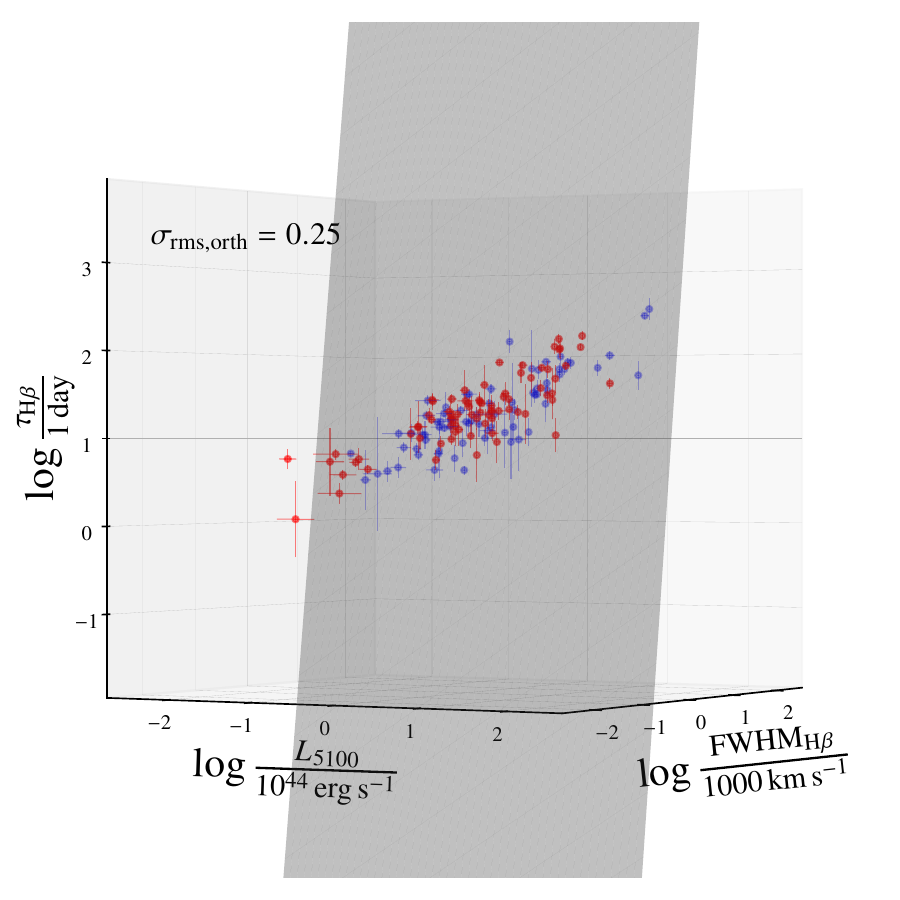}
\caption{The hyperplane (gray shaded area) defined by the best-fit relation from the conditional regression, assuming, \lag\ (left), \lumcont\ (middle), and \eddr\ (right) as the dependent variable, respectively. The best-fit relation with \eddr\ (Figure 13) is converted to the three-parameter relation with \vfwhm\ as described in Section 3.1. Note that the choice of the dependent variable significantly alters the resulting plane.}
\end{figure*}

\bibliography{bib.bib}

@ARTICLE{Akritas&Bershady96,
       author = {{Akritas}, Michael G. and {Bershady}, Matthew A.},
        title = "{Linear Regression for Astronomical Data with Measurement Errors and Intrinsic Scatter}",
      journal = {\apj},
         year = 1996,
        month = oct,
       volume = {470},
        pages = {706},
          doi = {10.1086/177901},
archivePrefix = {arXiv},
       eprint = {astro-ph/9605002},
 primaryClass = {astro-ph},
       adsurl = {https://ui.adsabs.harvard.edu/abs/1996ApJ...470..706A}
}

@article{Blandford+82,
       author = {{Blandford}, R.~D. and {McKee}, C.~F.},
        title = "{Reverberation mapping of the emission line regions of Seyfert galaxies and quasars.}",
      journal = {\apj},
         year = 1982,
        month = apr,
       volume = {255},
        pages = {419-439},
          doi = {10.1086/159843},
       adsurl = {https://ui.adsabs.harvard.edu/abs/1982ApJ...255..419B}
}

@article{Bahk+19,
	adsurl = {https://ui.adsabs.harvard.edu/abs/2019ApJ...875...50B},
	author = {{Bahk}, Hyeonguk and {Woo}, Jong-Hak and {Park}, Daeseong},
	doi = {10.3847/1538-4357/ab100d},
	eid = {50},
	journal = {\apj},
	month = apr,
	number = {1},
	pages = {50},
	title = {{Calibrating Mg II-based Black Hole Mass Estimators with H{\ensuremath{\beta}} Reverberation Measurements}},
	volume = {875},
	year = 2019}

@article{Bentz+09,
	adsurl = {https://ui.adsabs.harvard.edu/abs/2009ApJ...697..160B},
	archiveprefix = {arXiv},
	author = {{Bentz}, Misty C. and {Peterson}, Bradley M. and {Netzer}, Hagai and {Pogge}, Richard W. and {Vestergaard}, Marianne},
	doi = {10.1088/0004-637X/697/1/160},
	eprint = {0812.2283},
	journal = {\apj},
	month = may,
	number = {1},
	pages = {160-181},
	primaryclass = {astro-ph},
	title = {{The Radius-Luminosity Relationship for Active Galactic Nuclei: The Effect of Host-Galaxy Starlight on Luminosity Measurements. II. The Full Sample of Reverberation-Mapped AGNs}},
	volume = {697},
	year = 2009}

@article{Brotherton+15,
	adsurl = {https://ui.adsabs.harvard.edu/abs/2015MNRAS.451.1290B},
	archiveprefix = {arXiv},
	author = {{Brotherton}, Michael S. and {Runnoe}, J.~C. and {Shang}, Zhaohui and {DiPompeo}, M.~A.},
	doi = {10.1093/mnras/stv767},
	eprint = {1504.03427},
	journal = {\mnras},
	month = aug,
	number = {2},
	pages = {1290-1298},
	primaryclass = {astro-ph.GA},
	title = {{Bias in C IV-based quasar black hole mass scaling relationships from reverberation mapped samples}},
	volume = {451},
	year = 2015}

@article{Czerny+19,
	author = {Czerny, Bo{\.z}ena and Wang, Jian-Min and Du, Pu and Hryniewicz, Krzysztof and Karas, Vladimir and Li, Yan-Rong and Panda, Swayamtrupta and Sniegowska, Marzena and Wildy, Conor and Yuan, Ye-Fei},
	doi = {10.3847/1538-4357/aaf396},
	journal = {The Astrophysical Journal},
	month = {jan},
	number = {2},
	pages = {84},
	publisher = {The American Astronomical Society},
	title = {Interpretation of Departure from the Broad-line Region Scaling in Active Galactic Nuclei},
	url = {https://dx.doi.org/10.3847/1538-4357/aaf396},
	volume = {870},
	year = {2019}}

@ARTICLE{Collin06,
       author = {{Collin}, S. and {Kawaguchi}, T. and {Peterson}, B.~M. and {Vestergaard}, M.},
        title = "{Systematic effects in measurement of black hole masses by emission-line reverberation of active galactic nuclei: Eddington ratio and inclination}",
      journal = {\aap},
         year = 2006,
        month = sep,
       volume = {456},
       number = {1},
        pages = {75-90},
          doi = {10.1051/0004-6361:20064878},
archivePrefix = {arXiv},
       eprint = {astro-ph/0603460},
 primaryClass = {astro-ph},
       adsurl = {https://ui.adsabs.harvard.edu/abs/2006A&A...456...75C}
}

@article{Czerny+13,
  author = {Czerny, B. and Hryniewicz, K. and Maity, I. and Schwarzenberg-Czerny, A.},
  title = {Constraints on the Broad Line Region from Reverberation Mapping},
  journal = {A\&A},
  year = {2013},
  volume = {556},
  pages = {A97},
  doi = {10.1051/0004-6361/201220812}
}

@article{Gravity+24,
	adsurl = {https://ui.adsabs.harvard.edu/abs/2024A&A...684A.167G},
	archiveprefix = {arXiv},
	author = {{GRAVITY Collaboration} and {Amorim}, A. and {Bourdarot}, G. and {Brandner}, W. and {Cao}, Y. and {Cl{\'e}net}, Y. and {Davies}, R. and {de Zeeuw}, P.~T. and {Dexter}, J. and {Drescher}, A. and {Eckart}, A. and {Eisenhauer}, F. and {Fabricius}, M. and {Feuchtgruber}, H. and {F{\"o}rster Schreiber}, N.~M. and {Garcia}, P.~J.~V. and {Genzel}, R. and {Gillessen}, S. and {Gratadour}, D. and {H{\"o}nig}, S. and {Kishimoto}, M. and {Lacour}, S. and {Lutz}, D. and {Millour}, F. and {Netzer}, H. and {Ott}, T. and {Paumard}, T. and {Perraut}, K. and {Perrin}, G. and {Peterson}, B.~M. and {Petrucci}, P.~O. and {Pfuhl}, O. and {Prieto}, M.~A. and {Rabien}, S. and {Rouan}, D. and {Santos}, D.~J.~D. and {Shangguan}, J. and {Shimizu}, T. and {Sternberg}, A. and {Straubmeier}, C. and {Sturm}, E. and {Tacconi}, L.~J. and {Tristram}, K.~R.~W. and {Widmann}, F. and {Woillez}, J.},
	doi = {10.1051/0004-6361/202348167},
	eid = {A167},
	eprint = {2401.07676},
	journal = {\aap},
	month = apr,
	pages = {A167},
	primaryclass = {astro-ph.GA},
	title = {{The size-luminosity relation of local active galactic nuclei from interferometric observations of the broad-line region}},
	volume = {684},
	year = 2024}

@article{Grier+17,
	adsurl = {https://ui.adsabs.harvard.edu/abs/2017ApJ...851...21G},
	archiveprefix = {arXiv},
	author = {{Grier}, C.~J. and {Trump}, J.~R. and {Shen}, Yue and {Horne}, Keith and {Kinemuchi}, Karen and {McGreer}, Ian D. and {Starkey}, D.~A. and {Brandt}, W.~N. and {Hall}, P.~B. and {Kochanek}, C.~S. and {Chen}, Yuguang and {Denney}, K.~D. and {Greene}, Jenny E. and {Ho}, L.~C. and {Homayouni}, Y. and {I-Hsiu Li}, Jennifer and {Pei}, Liuyi and {Peterson}, B.~M. and {Petitjean}, P. and {Schneider}, D.~P. and {Sun}, Mouyuan and {AlSayyad}, Yusura and {Bizyaev}, Dmitry and {Brinkmann}, Jonathan and {Brownstein}, Joel R. and {Bundy}, Kevin and {Dawson}, K.~S. and {Eftekharzadeh}, Sarah and {Fernandez-Trincado}, J.~G. and {Gao}, Yang and {Hutchinson}, Timothy A. and {Jia}, Siyao and {Jiang}, Linhua and {Oravetz}, Daniel and {Pan}, Kaike and {Paris}, Isabelle and {Ponder}, Kara A. and {Peters}, Christina and {Rogerson}, Jesse and {Simmons}, Audrey and {Smith}, Robyn and {Wang}, Ran},
	doi = {10.3847/1538-4357/aa98dc},
	eid = {21},
	eprint = {1711.03114},
	journal = {\apj},
	month = dec,
	number = {1},
	pages = {21},
	primaryclass = {astro-ph.GA},
	title = {{The Sloan Digital Sky Survey Reverberation Mapping Project: H{\ensuremath{\alpha}} and H{\ensuremath{\beta}} Reverberation Measurements from First-year Spectroscopy and Photometry}},
	volume = {851},
	year = 2017}

@article{Inayoshi20,
	adsurl = {https://ui.adsabs.harvard.edu/abs/2020ARA&A..58...27I},
	archiveprefix = {arXiv},
	author = {{Inayoshi}, Kohei and {Visbal}, Eli and {Haiman}, Zolt{\'a}n},
	doi = {10.1146/annurev-astro-120419-014455},
	eprint = {1911.05791},
	journal = {\araa},
	month = aug,
	pages = {27-97},
	primaryclass = {astro-ph.GA},
	title = {{The Assembly of the First Massive Black Holes}},
	volume = {58},
	year = 2020}

@article{Martinez+20,
	adsurl = {https://ui.adsabs.harvard.edu/abs/2020ApJ...903...86M},
	archiveprefix = {arXiv},
	author = {{Mart{\'\i}nez-Aldama}, Mary Loli and {Zaja{\v{c}}ek}, Michal and {Czerny}, Bo{\.z}ena and {Panda}, Swayamtrupta},
	doi = {10.3847/1538-4357/abb6f8},
	eid = {86},
	eprint = {2007.09955},
	journal = {\apj},
	month = nov,
	number = {2},
	pages = {86},
	primaryclass = {astro-ph.GA},
	title = {{Scatter Analysis along the Multidimensional Radius-Luminosity Relations for Reverberation-mapped Mg II Sources}},
	volume = {903},
	year = 2020}

@article{Naddaf+25,
	adsurl = {https://ui.adsabs.harvard.edu/abs/2025arXiv250601159N},
	archiveprefix = {arXiv},
	author = {{Naddaf}, M.~H. and {Martinez-Aldama}, M.~L. and {Hutsemekers}, D. and {Savic}, D. and {Czerny}, B.},
	doi = {10.48550/arXiv.2506.01159},
	eid = {arXiv:2506.01159},
	eprint = {2506.01159},
	journal = {arXiv e-prints},
	month = jun,
	pages = {arXiv:2506.01159},
	primaryclass = {astro-ph.GA},
	title = {{H$β$ line shape and radius-luminosity relation in 2.5D FRADO}},
	year = 2025}

@article{Du+15,
	adsurl = {https://ui.adsabs.harvard.edu/abs/2015ApJ...806...22D},
	archiveprefix = {arXiv},
	author = {{Du}, Pu and {Hu}, Chen and {Lu}, Kai-Xing and {Huang}, Ying-Ke and {Cheng}, Cheng and {Qiu}, Jie and {Li}, Yan-Rong and {Zhang}, Yang-Wei and {Fan}, Xu-Liang and {Bai}, Jin-Ming and {Bian}, Wei-Hao and {Yuan}, Ye-Fei and {Kaspi}, Shai and {Ho}, Luis C. and {Netzer}, Hagai and {Wang}, Jian-Min and {SEAMBH Collaboration}},
	doi = {10.1088/0004-637X/806/1/22},
	eid = {22},
	eprint = {1504.01844},
	journal = {\apj},
	month = jun,
	number = {1},
	pages = {22},
	primaryclass = {astro-ph.GA},
	title = {{Supermassive Black Holes with High Accretion Rates in Active Galactic Nuclei. IV. H{\ensuremath{\beta}} Time Lags and Implications for Super-Eddington Accretion}},
	volume = {806},
	year = 2015}

@article{Du+16,
	adsurl = {https://ui.adsabs.harvard.edu/abs/2016ApJ...825..126D},
	archiveprefix = {arXiv},
	author = {{Du}, Pu and {Lu}, Kai-Xing and {Zhang}, Zhi-Xiang and {Huang}, Ying-Ke and {Wang}, Kai and {Hu}, Chen and {Qiu}, Jie and {Li}, Yan-Rong and {Fan}, Xu-Liang and {Fang}, Xiang-Er and {Bai}, Jin-Ming and {Bian}, Wei-Hao and {Yuan}, Ye-Fei and {Ho}, Luis C. and {Wang}, Jian-Min and {SEAMBH Collaboration}},
	doi = {10.3847/0004-637X/825/2/126},
	eid = {126},
	eprint = {1604.06218},
	journal = {\apj},
	month = jul,
	number = {2},
	pages = {126},
	primaryclass = {astro-ph.GA},
	title = {{Supermassive Black Holes with High Accretion Rates in Active Galactic Nuclei. V. A New Size-Luminosity Scaling Relation for the Broad-line Region}},
	volume = {825},
	year = 2016}

@article{Du+18,
	author = {Du, Pu and Brotherton, Michael and Wang, Jian-Min and others},
	doi = {10.3847/1538-4357/aaae6b},
	journal = {The Astrophysical Journal},
	number = {1},
	pages = {6},
	title = {Exploring the Hβ Time Lags for Super-Eddington Accreting Massive Black Holes},
	volume = {856},
	year = {2018}}

@article{Du&Wang19,
	adsurl = {https://ui.adsabs.harvard.edu/abs/2019ApJ...886...42D},
	archiveprefix = {arXiv},
	author = {{Du}, Pu and {Wang}, Jian-Min},
	doi = {10.3847/1538-4357/ab4908},
	eid = {42},
	eprint = {1909.06735},
	journal = {\apj},
	month = nov,
	number = {1},
	pages = {42},
	primaryclass = {astro-ph.GA},
	title = {{The Radius-Luminosity Relationship Depends on Optical Spectra in Active Galactic Nuclei}},
	volume = {886},
	year = 2019}

@article{Kovasevic10,
	adsurl = {https://ui.adsabs.harvard.edu/abs/2010ApJS..189...15K},
	archiveprefix = {arXiv},
	author = {{Kova{\v{c}}evi{\'c}}, Jelena and {Popovi{\'c}}, Luka {\v{C}}. and {Dimitrijevi{\'c}}, Milan S.},
	doi = {10.1088/0067-0049/189/1/15},
	eprint = {1004.2212},
	journal = {\apjs},
	month = jul,
	number = {1},
	pages = {15-36},
	primaryclass = {astro-ph.CO},
	title = {{Analysis of Optical Fe II Emission in a Sample of Active Galactic Nucleus Spectra}},
	volume = {189},
	year = 2010}

@article{Li+21,
	author = {Li, Sha-Sha and Yang, Sen and Yang, Zi-Xu and Chen, Yong-Jie and Songsheng, Yu-Yang and Liu, He-Zhen and Du, Pu and Luo, Bin and Yu, Zhe and Hu, Chen and Jiang, Bo-Wei and Bao, Dong-Wei and Guo, Wei-Jian and Zhang, Zhi-Xiang and Li, Yan-Rong and Xiao, Ming and Lu, Kai-Xing and Ho, Luis C. and Bai, Jin-Ming and Bian, Wei-Hao and Aceituno, Jes{\'u}s and Minezaki, Takeo and Horne, Keith and Kokubo, Mitsuru and Wang, Jian-Min},
	doi = {10.3847/1538-4357/ac116e},
	journal = {The Astrophysical Journal},
	month = {oct},
	number = {1},
	pages = {9},
	publisher = {The American Astronomical Society},
	title = {Reverberation Mapping of Two Luminous Quasars: The Broad-line Region Structure and Black Hole Mass},
	url = {https://dx.doi.org/10.3847/1538-4357/ac116e},
	volume = {920},
	year = {2021}}

@article{Maithil+22,
	author = {Maithil, Jaya and Brotherton, Michael S and Shemmer, Ohad and Du, Pu and Wang, Jian-Min and Myers, Adam D and McLane, Jacob N and Dix, Cooper and Matthews, Brandon M},
	doi = {10.1093/mnras/stac1748},
	eprint = {https://academic.oup.com/mnras/article-pdf/515/1/491/44926916/stac1748.pdf},
	issn = {0035-8711},
	journal = {Monthly Notices of the Royal Astronomical Society},
	month = {06},
	number = {1},
	pages = {491-506},
	title = {Systematically smaller single-epoch quasar black hole masses using a radius--luminosity relationship corrected for spectral bias},
	url = {https://doi.org/10.1093/mnras/stac1748},
	volume = {515},
	year = {2022}}

@article{MartinezAldama19,
  author = {Mart{\'i}nez-Aldama, M. L. and Czerny, B. and Du, P. and et al.},
  title = {Calibration of Super-Eddington AGNs as Cosmological Probes},
  journal = {ApJ},
  year = {2019},
  volume = {883},
  pages = {170},
  doi = {10.3847/1538-4357/ab3bfe}
}

@article{Marziani03,
	adsurl = {https://ui.adsabs.harvard.edu/abs/2003ApJS..145..199M},
	author = {{Marziani}, P. and {Sulentic}, J.~W. and {Zamanov}, R. and {Calvani}, M. and {Dultzin-Hacyan}, D. and {Bachev}, R. and {Zwitter}, T.},
	doi = {10.1086/346025},
	journal = {\apjs},
	month = apr,
	number = {2},
	pages = {199-211},
	title = {{An Optical Spectroscopic Atlas of Low-Redshift Active Galactic Nuclei}},
	volume = {145},
	year = 2003}

@article{Wang+14,
	author = {Wang, Jian-Min and Du, Pu and Hu, Chen and others},
	doi = {10.1088/0004-637X/793/2/108},
	journal = {The Astrophysical Journal},
	number = {2},
	pages = {108},
	title = {Supermassive Black Holes with High Accretion Rates in Active Galactic Nuclei. I. First Results from a New Reverberation Mapping Campaign},
	volume = {793},
	year = {2014}}

@article{Kaspi+00,
	author = {Kaspi, Shai and Smith, Paul S. and Netzer, Hagai and Maoz, Dan and Jannuzi, Buell T. and Giveon, Uriel},
	doi = {10.1086/308704},
	journal = {The Astrophysical Journal},
	month = {apr},
	number = {2},
	pages = {631},
	title = {Reverberation Measurements for 17 Quasars and the Size-Mass-Luminosity Relations in Active Galactic Nuclei},
	url = {https://dx.doi.org/10.1086/308704},
	volume = {533},
	year = {2000}}

@article{Pan+25,
	adsurl = {https://ui.adsabs.harvard.edu/abs/2025ApJ...987...48P},
	archiveprefix = {arXiv},
	author = {{Pan}, Zhiwei and {Jiang}, Linhua and {Guo}, Wei-Jian and {Sun}, Shengxiu and {Siudek}, Ma{\l}gorzata and {Aguilar}, Jessica Nicole and {Ahlen}, Steven and {Brooks}, David and {Claybaugh}, Todd and {de la Macorra}, Axel and {Doel}, Peter and {Gazta{\~n}aga}, Enrique and {Gontcho A Gontcho}, Satya and {Juneau}, Stephanie and {Kisner}, Theodore and {Lambert}, Andrew and {Landriau}, Martin and {Le Guillou}, Laurent and {Manera}, Marc and {Martini}, Paul and {Meisner}, Aaron and {Miquel}, Ramon and {Moustakas}, John and {Myers}, Adam and {Poppett}, Claire and {Prada}, Francisco and {Rossi}, Graziano and {Sanchez}, Eusebio and {Schubnell}, Michael and {Seo}, Hee-Jong and {Sprayberry}, David and {Tarl{\'e}}, Gregory and {Weaver}, Benjamin Alan and {Zou}, Hu},
	doi = {10.3847/1538-4357/add7dd},
	eid = {48},
	eprint = {2502.03684},
	journal = {\apj},
	month = jul,
	number = {1},
	pages = {48},
	primaryclass = {astro-ph.GA},
	title = {{Iron-corrected Single-epoch Black Hole Masses of DESI Quasars at Low Redshift}},
	volume = {987},
	year = 2025}

@article{Park+12,
	adsurl = {https://ui.adsabs.harvard.edu/abs/2012ApJS..203....6P},
	archiveprefix = {arXiv},
	author = {{Park}, Daeseong and {Kelly}, Brandon C. and {Woo}, Jong-Hak and {Treu}, Tommaso},
	doi = {10.1088/0067-0049/203/1/6},
	eid = {6},
	eprint = {1209.3773},
	journal = {\apjs},
	month = nov,
	number = {1},
	pages = {6},
	primaryclass = {astro-ph.CO},
	title = {{Recalibration of the Virial Factor and M $_{BH}$-{\ensuremath{\sigma}}$_{*}$ Relation for Local Active Galaxies}},
	volume = {203},
	year = 2012}

@article{Park22,
	adsurl = {https://ui.adsabs.harvard.edu/abs/2022ApJS..258...38P},
	archiveprefix = {arXiv},
	author = {{Park}, Daeseong and {Barth}, Aaron J. and {Ho}, Luis C. and {Laor}, Ari},
	doi = {10.3847/1538-4365/ac3f3e},
	eid = {38},
	eprint = {2111.15118},
	journal = {\apjs},
	month = feb,
	number = {2},
	pages = {38},
	primaryclass = {astro-ph.GA},
	title = {{A New Iron Emission Template for Active Galactic Nuclei. I. Optical Template for the H{\ensuremath{\beta}} Region}},
	volume = {258},
	year = 2022}

@article{Shen+15,
	adsurl = {https://ui.adsabs.harvard.edu/abs/2015ApJS..216....4S},
	archiveprefix = {arXiv},
	author = {{Shen}, Yue and {Brandt}, W.~N. and {Dawson}, Kyle S. and {Hall}, Patrick B. and {McGreer}, Ian D. and {Anderson}, Scott F. and {Chen}, Yuguang and {Denney}, Kelly D. and {Eftekharzadeh}, Sarah and {Fan}, Xiaohui and {Gao}, Yang and {Green}, Paul J. and {Greene}, Jenny E. and {Ho}, Luis C. and {Horne}, Keith and {Jiang}, Linhua and {Kelly}, Brandon C. and {Kinemuchi}, Karen and {Kochanek}, Christopher S. and {P{\^a}ris}, Isabelle and {Peters}, Christina M. and {Peterson}, Bradley M. and {Petitjean}, Patrick and {Ponder}, Kara and {Richards}, Gordon T. and {Schneider}, Donald P. and {Seth}, Anil and {Smith}, Robyn N. and {Strauss}, Michael A. and {Tao}, Charling and {Trump}, Jonathan R. and {Wood-Vasey}, W.~M. and {Zu}, Ying and {Eisenstein}, Daniel J. and {Pan}, Kaike and {Bizyaev}, Dmitry and {Malanushenko}, Viktor and {Malanushenko}, Elena and {Oravetz}, Daniel},
	doi = {10.1088/0067-0049/216/1/4},
	eid = {4},
	eprint = {1408.5970},
	journal = {\apjs},
	month = jan,
	number = {1},
	pages = {4},
	primaryclass = {astro-ph.IM},
	title = {{The Sloan Digital Sky Survey Reverberation Mapping Project: Technical Overview}},
	volume = {216},
	year = 2015}

@article{Vestergaard02,
  author  = {Vestergaard, Marianne},
  title   = {Determining Central Black Hole Masses in Distant Active Galaxies and Quasars. II. Improved Optical and UV Scaling Relationships},
  journal = {ApJ},
  year    = {2002},
  volume  = {571},
  pages   = {733--752},
  doi     = {10.1086/340102}
}

@ARTICLE{Villafana+23,
       author = {{Villafa{\~n}a}, Lizvette and {Williams}, Peter R. and {Treu}, Tommaso and {Brewer}, Brendon J. and {Barth}, Aaron J. and {U}, Vivian and {Bennert}, Vardha N. and {Guo}, Hengxiao and {Bentz}, Misty C. and {Canalizo}, Gabriela and {Filippenko}, Alexei V. and {Gates}, Elinor and {Joner}, Michael D. and {Malkan}, Matthew A. and {Woo}, Jong-Hak and {Abolfathi}, Bela and {Bohn}, Thomas and {Bostroem}, K. Azalee and {Brandel}, Andrew and {Brink}, Thomas G. and {Channa}, Sanyum and {Cosens}, Maren and {Donohue}, Edward and {Halevi}, Goni and {Hood}, Carol E. and {Horst}, J. Chuck and {de Kouchkovsky}, Maxime and {Kuhn}, Benjamin and {Leonard}, Douglas C. and {Michel}, Ra{\'u}l and {B. Olaes}, Melanie Kae and {Park}, Daeseong and {Runco}, Jordan N. and {Sexton}, Remington O. and {Shivvers}, Isaac and {Spencer}, Chance L. and {Stahl}, Benjamin E. and {Stegman}, Samantha and {Walsh}, Jonelle L. and {Zheng}, WeiKang},
        title = "{What Does the Geometry of the H{\ensuremath{\beta}} BLR Depend On?}",
      journal = {\apj},
         year = 2023,
        month = may,
       volume = {948},
       number = {2},
          eid = {95},
        pages = {95},
          doi = {10.3847/1538-4357/accc84},
       adsurl = {https://ui.adsabs.harvard.edu/abs/2023ApJ...948...95V}
}

@article{Watson+11,
  author = {Watson, D. and Denney, K. D. and Vestergaard, M. and Davis, T. M.},
  title = {A New Cosmological Distance Measure Using Active Galactic Nuclei},
  journal = {ApJ},
  year = {2011},
  volume = {740},
  pages = {L49},
  doi = {10.1088/2041-8205/740/2/L49}
}

@article{WooUrry02,
  author  = {Woo, Jong-Hak and Urry, C. Megan},
  title   = {Active Galactic Nucleus Black Hole Masses and Bolometric Luminosities},
  journal = {ApJ},
  year    = {2002},
  volume  = {579},
  pages   = {530--544},
  doi     = {10.1086/342878}
}

@article{Woo+13,
	adsurl = {https://ui.adsabs.harvard.edu/abs/2013ApJ...772...49W},
	archiveprefix = {arXiv},
	author = {{Woo}, Jong-Hak and {Schulze}, Andreas and {Park}, Daeseong and {Kang}, Wol-Rang and {Kim}, Sang Chul and {Riechers}, Dominik A.},
	doi = {10.1088/0004-637X/772/1/49},
	eid = {49},
	eprint = {1305.2946},
	journal = {\apj},
	month = jul,
	number = {1},
	pages = {49},
	primaryclass = {astro-ph.CO},
	title = {{Do Quiescent and Active Galaxies Have Different M $_{BH}$-{\ensuremath{\sigma}}$_{*}$ Relations?}},
	volume = {772},
	year = 2013}

@article{Wang&Woo24,
       author = {{Wang}, Shu and {Woo}, Jong-Hak},
        title = "{Revisiting the H{\ensuremath{\beta}} Size{\textendash}Luminosity Relation Using a Uniform Reverberation-mapping Analysis}",
      journal = {\apjs},
         year = 2024,
        month = nov,
       volume = {275},
       number = {1},
          eid = {13},
        pages = {13},
          doi = {10.3847/1538-4365/ad74f2},
archivePrefix = {arXiv},
       eprint = {2408.15872},
 primaryClass = {astro-ph.GA},
       adsurl = {https://ui.adsabs.harvard.edu/abs/2024ApJS..275...13W}
}

@article{Woo+15,
	adsurl = {https://ui.adsabs.harvard.edu/\#abs/2015ApJ...801...38W},
	archiveprefix = {arXiv},
	author = {{Woo}, Jong-Hak and {Yoon}, Yosep and {Park}, Songyoun and {Park}, Daeseong and {Kim}, Sang Chul},
	doi = {10.1088/0004-637X/801/1/38},
	eid = {38},
	eprint = {1412.7225},
	journal = {\apj},
	month = Mar,
	pages = {38},
	primaryclass = {astro-ph.GA},
	title = {{The Black Hole Mass-Stellar Velocity Dispersion Relation of Narrow-line Seyfert 1 Galaxies}},
	volume = {801},
	year = {2015}}

@article{Woo+24,
	adsurl = {https://ui.adsabs.harvard.edu/abs/2024ApJ...962...67W},
	archiveprefix = {arXiv},
	author = {{Woo}, Jong-Hak and {Wang}, Shu and {Rakshit}, Suvendu and {Cho}, Hojin and {Son}, Donghoon and {Bennert}, Vardha N. and {Gallo}, Elena and {Hodges-Kluck}, Edmund and {Treu}, Tommaso and {Barth}, Aaron J. and {Cho}, Wanjin and {Foord}, Adi and {Geum}, Jaehyuk and {Guo}, Hengxiao and {Jadhav}, Yashashree and {Jeon}, Yiseul and {Kabasares}, Kyle M. and {Kang}, Won-Suk and {Kim}, Changseok and {Kim}, Minjin and {Kim}, Tae-Woo and {Le}, Huynh Anh N. and {Malkan}, Matthew A. and {Mandal}, Amit Kumar and {Park}, Daeseong and {Spencer}, Chance and {Shin}, Jaejin and {Sung}, Hyun-il and {U}, Vivian and {Williams}, Peter R. and {Yee}, Nick},
	doi = {10.3847/1538-4357/ad132f},
	eid = {67},
	eprint = {2311.15518},
	journal = {\apj},
	month = feb,
	number = {1},
	pages = {67},
	primaryclass = {astro-ph.GA},
	title = {{The Seoul National University AGN Monitoring Project. III. H{\ensuremath{\beta}} Lag Measurements of 32 Luminous Active Galactic Nuclei and the High-luminosity End of the Size{\textendash}Luminosity Relation}},
	volume = {962},
	year = 2024}

@article{HyperFit,
	adsurl = {https://ui.adsabs.harvard.edu/abs/2015PASA...32...33R},
	archiveprefix = {arXiv},
	author = {{Robotham}, A.~S.~G. and {Obreschkow}, D.},
	doi = {10.1017/pasa.2015.33},
	eid = {e033},
	eprint = {1508.02145},
	journal = {\pasa},
	month = sep,
	pages = {e033},
	primaryclass = {astro-ph.IM},
	title = {{Hyper-Fit: Fitting Linear Models to Multidimensional Data with Multivariate Gaussian Uncertainties}},
	volume = {32},
	year = 2015}

@article{Woo&Urry02,
	adsurl = {https://ui.adsabs.harvard.edu/abs/2002ApJ...579..530W},
	author = {{Woo}, J.-H. and {Urry}, C.~M.},
	doi = {10.1086/342878},
	eprint = {astro-ph/0207249},
	journal = {\apj},
	month = nov,
	pages = {530-544},
	title = {{Active Galactic Nucleus Black Hole Masses and Bolometric Luminosities}},
	volume = 579,
	year = 2002}

@article{Bentz+13,
	adsurl = {https://ui.adsabs.harvard.edu/\#abs/2013ApJ...767..149B},
	archiveprefix = {arXiv},
	author = {{Bentz}, Misty C. and {Denney}, Kelly D. and {Grier}, Catherine J. and {Barth}, Aaron J. and {Peterson}, Bradley M. and {Vestergaard}, Marianne and {Bennert}, Vardha N. and {Canalizo}, Gabriela and {De Rosa}, Gisella and {Filippenko}, Alexei V. and {Gates}, Elinor L. and {Greene}, Jenny E. and {Li}, Weidong and {Malkan}, Matthew A. and {Pogge}, Richard W. and {Stern}, Daniel and {Treu}, Tommaso and {Woo}, Jong-Hak},
	doi = {10.1088/0004-637X/767/2/149},
	eid = {149},
	eprint = {1303.1742},
	journal = {\apj},
	month = Apr,
	pages = {149},
	primaryclass = {astro-ph.CO},
	title = {{The Low-luminosity End of the Radius-Luminosity Relationship for Active Galactic Nuclei}},
	volume = {767},
	year = {2013}}

@article{Rakshit+20,
	author = {Rakshit, Suvendu and Stalin, C. S. and Kotilainen, Jari},
	doi = {10.3847/1538-4365/ab99c5},
	journal = {The Astrophysical Journal Supplement Series},
	month = {jul},
	number = {1},
	pages = {17},
	publisher = {The American Astronomical Society},
	title = {Spectral Properties of Quasars from Sloan Digital Sky Survey Data Release 14: The Catalog},
	url = {https://dx.doi.org/10.3847/1538-4365/ab99c5},
	volume = {249},
	year = {2020}}

@article{Le+20,
	adsurl = {https://ui.adsabs.harvard.edu/abs/2020ApJ...901...35L},
	archiveprefix = {arXiv},
	author = {{Le}, Huynh Anh N. and {Woo}, Jong-Hak and {Xue}, Yongquan},
	doi = {10.3847/1538-4357/abada0},
	eid = {35},
	eprint = {2008.02990},
	journal = {\apj},
	month = sep,
	number = {1},
	pages = {35},
	primaryclass = {astro-ph.GA},
	title = {{Calibrating Mg II-based Black Hole Mass Estimators Using Low-to-high-luminosity Active Galactic Nuclei}},
	volume = {901},
	year = 2020}

@article{Wang+09,
	adsurl = {https://ui.adsabs.harvard.edu/abs/2009ApJ...707.1334W},
	archiveprefix = {arXiv},
	author = {{Wang}, Jian-Guo and {Dong}, Xiao-Bo and {Wang}, Ting-Gui and {Ho}, Luis C. and {Yuan}, Weimin and {Wang}, Huiyuan and {Zhang}, Kai and {Zhang}, Shaohua and {Zhou}, Hongyan},
	doi = {10.1088/0004-637X/707/2/1334},
	eprint = {0910.2848},
	journal = {\apj},
	month = dec,
	number = {2},
	pages = {1334-1346},
	primaryclass = {astro-ph.CO},
	title = {{Estimating Black Hole Masses in Active Galactic Nuclei Using the Mg II {\ensuremath{\lambda}}2800 Emission Line}},
	volume = {707},
	year = 2009}

@article{Pancoast+14,
	adsurl = {https://ui.adsabs.harvard.edu/abs/2014MNRAS.445.3073P},
	archiveprefix = {arXiv},
	author = {{Pancoast}, Anna and {Brewer}, Brendon J. and {Treu}, Tommaso and {Park}, Daeseong and {Barth}, Aaron J. and {Bentz}, Misty C. and {Woo}, Jong-Hak},
	doi = {10.1093/mnras/stu1419},
	eprint = {1311.6475},
	journal = {\mnras},
	month = dec,
	number = {3},
	pages = {3073-3091},
	primaryclass = {astro-ph.CO},
	title = {{Modelling reverberation mapping data - II. Dynamical modelling of the Lick AGN Monitoring Project 2008 data set}},
	volume = {445},
	year = 2014}

@article{Peterson+93,
       author = {{Peterson}, Bradley M.},
        title = "{Reverberation Mapping of Active Galactic Nuclei}",
      journal = {\pasp},
         year = 1993,
        month = mar,
       volume = {105},
        pages = {247},
          doi = {10.1086/133140},
       adsurl = {https://ui.adsabs.harvard.edu/abs/1993PASP..105..247P}
}

@article{Wang+19,
	adsurl = {https://ui.adsabs.harvard.edu/abs/2019ApJ...882....4W},
	archiveprefix = {arXiv},
	author = {{Wang}, Shu and {Shen}, Yue and {Jiang}, Linhua and {Horne}, Keith and {Brandt}, W.~N. and {Grier}, C.~J. and {Ho}, Luis C. and {Homayouni}, Yasaman and {I-Hsiu Li}, Jennifer and {Schneider}, Donald P. and {Trump}, Jonathan R.},
	doi = {10.3847/1538-4357/ab322b},
	eid = {4},
	eprint = {1903.10015},
	journal = {\apj},
	month = sep,
	number = {1},
	pages = {4},
	primaryclass = {astro-ph.GA},
	title = {{The Sloan Digital Sky Survey Reverberation Mapping Project: Low-ionization Broad-line Widths and Implications for Virial Black Hole Mass Estimation}},
	volume = {882},
	year = 2019}

@article{Williams+18,
	adsurl = {https://ui.adsabs.harvard.edu/abs/2018ApJ...866...75W},
	archiveprefix = {arXiv},
	author = {{Williams}, Peter R. and {Pancoast}, Anna and {Treu}, Tommaso and {Brewer}, Brendon J. and {Barth}, Aaron J. and {Bennert}, Vardha N. and {Buehler}, Tabitha and {Canalizo}, Gabriela and {Cenko}, S. Bradley and {Clubb}, Kelsey I. and {Cooper}, Michael C. and {Filippenko}, Alexei V. and {Gates}, Elinor and {Hoenig}, Sebastian F. and {Joner}, Michael D. and {Kandrashoff}, Michael T. and {Laney}, Clifton David and {Lazarova}, Mariana S. and {Li}, Weidong and {Malkan}, Matthew A. and {Rex}, Jacob and {Silverman}, Jeffrey M. and {Tollerud}, Erik and {Walsh}, Jonelle L. and {Woo}, Jong-Hak},
	doi = {10.3847/1538-4357/aae086},
	eid = {75},
	eprint = {1809.05113},
	journal = {\apj},
	month = oct,
	number = {2},
	pages = {75},
	primaryclass = {astro-ph.GA},
	title = {{The Lick AGN Monitoring Project 2011: Dynamical Modeling of the Broad-line Region}},
	volume = {866},
	year = 2018}

@article{Wu+25,
	adsurl = {https://ui.adsabs.harvard.edu/abs/2025ApJ...980..134W},
	archiveprefix = {arXiv},
	author = {{Wu}, Qiaoya and {Shen}, Yue and {Guo}, Hengxiao and {Anderson}, Scott F. and {Brandt}, W.~N. and {Grier}, Catherine J. and {Hall}, Patrick B. and {Ho}, Luis C. and {Homayouni}, Yasaman and {Horne}, Keith and {Li}, Jennifer I. -Hsiu and {Schneider}, Donald P.},
	doi = {10.3847/1538-4357/ada386},
	eid = {134},
	eprint = {2407.01737},
	journal = {\apj},
	month = feb,
	number = {1},
	pages = {134},
	primaryclass = {astro-ph.GA},
	title = {{Understanding the Broad-line Region of Active Galactic Nuclei with Photoionization. I. The Moderate-accretion Regime}},
	volume = {980},
	year = 2025}

@article{Yu+23,
  author = {Yu, Liang and Wang, Jian-Min and Du, Pu and et al.},
  title = {Reverberation-mapped Active Galactic Nuclei as Cosmological Probes: An Extended Hubble Diagram},
  journal = {ApJ},
  year = {2023},
  volume = {954},
  pages = {26},
  doi = {10.3847/1538-4357/acdd5e}
}

@ARTICLE{Isobe+90,
       author = {{Isobe+90}, Takashi and {Feigelson}, Eric D. and {Akritas}, Michael G. and {Babu}, Gutti Jogesh},
        title = "{Linear Regression in Astronomy. I.}",
      journal = {\apj},
         year = 1990,
        month = nov,
       volume = {364},
        pages = {104},
          doi = {10.1086/169390},
       adsurl = {https://ui.adsabs.harvard.edu/abs/1990ApJ...364..104I}
}

\end{document}